\documentclass[conference]{IEEEtran}
\IEEEoverridecommandlockouts
\usepackage{cite}
\usepackage{subcaption}

\usepackage{float}
\usepackage{amsmath,amssymb,amsfonts}
\usepackage{mathtools}
\usepackage{amsthm}

\usepackage{graphicx}
\usepackage{booktabs}
\usepackage{multirow}
\usepackage{array}
\usepackage{adjustbox}
\usepackage{diagbox}
\usepackage{makecell}
\usepackage[table]{xcolor}

\usepackage[algo2e,ruled,vlined,linesnumbered]{algorithm2e}

\usepackage{rotating}
\usepackage{pdflscape}

\usepackage{textcomp}

\usepackage[hyphens]{url}
\usepackage[hidelinks]{hyperref}
\usepackage{xcolor}
\def\BibTeX{{\rm B\kern-.05em{\sc i\kern-.025em b}\kern-.08em
    T\kern-.1667em\lower.7ex\hbox{E}\kern-.125emX}}

    \usepackage{amsthm}   % provides theorem environments
\theoremstyle{plain}
\newtheorem{proposition}{Proposition}

\begin{document}
\raggedbottom

\author{
\IEEEauthorblockN{
Ynes Ineza, Gerald Jackson, Prince Niyonkuru, Jaden Kevil, and Abdul Serwadda
}
\IEEEauthorblockA{
Department of Computer Science, Texas Tech University, Lubbock, TX, USA\\
Email: \{yineza, gerjacks, pniyonku, jadkevil\}@ttu.edu,
abdul.serwadda@ttu.edu
}

}

\title{Intermittent File Encryption in Ransomware: Measurement, Modeling, and Detection}

\maketitle

\begin{abstract}
 File-encrypting ransomware increasingly employs intermittent encryption techniques, encrypting only parts of files to evade classical detection methods.This paper provides a systematic empirical characterization of byte-level statistics under intermittent encryption across common file types, establishing a baseline for how partial encryption reshapes data structure. 
 
 Guided by these measurements, we model intermittent encryption as a convex mixture of ciphertext and cleartext and, via a classical KL-divergence bound, derive file-type–specific detectability limits for histogram-based detectors. Leveraging these insights, we evaluate convolutional neural network (CNN) detectors trained on realistic intermittent-encryption configurations from leading ransomware families. Our findings show that localized, chunk-level CNNs consistently outperform whole-file analysis, highlighting a practical, robust baseline for future detection systems.
\end{abstract}


\section{Introduction}
Over the past decade, file-encrypting ransomware has emerged as one of the most prevalent and damaging cyber threats. Recent industry reports highlight a dramatic rise in ransomware incidents, with total infections increasing by over 200\% from 2023 to 2024, according to Microsoft’s Digital Defense Report \cite{microsoft2023}. This surge is partly driven by the proliferation of Ransomware-as-a-Service (RaaS) platforms, significantly lowering entry barriers for cybercriminals. Instead of developing specialized malware from scratch, threat actors now commonly lease established ransomware toolkits, configure them for targeted attacks, and share ransom profits with the developers \cite{microsoft2023}. %Notably, high-profile ransomware families such as BlackCat (ALPHV) \cite{picus2025} exemplify this trend by providing customizable \emph{encryption configurations}, further cementing ransomware’s dominance as a major security concern.

Defensively, research has proposed various mechanisms for detecting and mitigating file-encrypting ransomware, broadly classified into three approaches: (1) monitoring API or system calls to identify suspicious patterns of cryptographic library invocations or unusual system write behaviors; (2) observing input/output (I/O) requests to detect large-scale file modifications indicative of ransomware activity; and (3) analyzing file-system content directly by measuring byte-level characteristics such as entropy and statistical randomness. The third approach capitalizes on the fact that fully encrypted files typically exhibit high statistical randomness, making them identifiable through entropy-based methods.

This paper focuses specifically on content-based detection, recognizing that recent ransomware variants intentionally evade file-structure detection techniques by employing intermittent encryption strategies. Rather than encrypting entire files outright, these newer malware families encrypt only selected portions of files. BlackCat ransomware, for example, offers multiple intermittent encryption modes such as “dot pattern,” “smart pattern,” and “head only,” each configurable with parameters defining the encrypted segments' size and spacing \cite{cyberark2025}. Consequently, attackers can carefully manipulate a file’s randomness profile, causing it to diverge significantly from the typical signatures of fully encrypted data and evade entropy-based detection methods.

The challenge deepens with the inherent diversity across file formats. Different file types exhibit distinct baseline randomness patterns: uncompressed formats like XLS files present relatively low entropy, losslessly compressed formats such as PNG have moderate entropy, and formats with lossy compression, including MP4 files, generally display higher baseline randomness. When sophisticated intermittent encryption techniques intersect with these diverse baseline structures, the resulting data signatures can appear deceptively benign, further complicating effective detection.

In this paper, we systematically examine the low-level file structure dynamics underlying intermittent encryption schemes before leveraging these insights to develop robust detection methods. %We begin by empirically characterizing how byte-level statistics evolve under varying degrees of intermittent encryption across common file types, providing critical baseline insights into how partial encryption impacts file structure. Next, we introduce a theoretical framework to establish clear and rigorous boundaries for the effectiveness of entropy-based detection, considering inherent differences across file formats. Finally, we evaluate practical detection strategies by implementing and rigorously testing machine learning-based detection pipelines trained explicitly on realistic ransomware configurations. 

Specifically, our study makes \textbf{three primary contributions}:

\textbf{(1) Systematic, multi-type empirical atlas of partial-encryption artefacts.}
Prior work in cryptography and digital forensics largely contrasts the byte-level structure of unencrypted files vs. fully encrypted endpoints. We provide a \textit{systematic, multi-file-type characterization of how byte-level structure evolves continuously with encryption coverage} ($0$–$100\%$) under ransomware-inspired patterns. %Across eleven representative formats, we report dense coverage curves with quartile/decile bands and apply Mann–Kendall and Sen’s slope to establish monotone trends and effect sizes for entropy, variance, and skewness. 
The atlas exposes file-type–specific rates of change and directly motivates file-type-aware thresholds and localized (chunk-level) analysis where global statistics are weak.

\textbf{(2) Mixture-based analytic model with file-type detectability ceilings.}
We formalize partial encryption as a blend of ciphertext with a file’s native byte histogram and analyze how the observable distribution drifts as coverage increases, yielding an “escape trajectory’’ that links coverage to the disappearance of recognizable structure. Building on this, we tailor a standard information-theoretic bound to the intermittent-encryption mixture, which produces a single, file-type–specific constant that captures how quickly structure vanishes. We estimate this constant directly from cleartext corpora, and use it to compute practical detectability ceilings for entropy- and histogram-based detectors, to set family-aware thresholds that indicate up to what coverage alerts can still fire reliably.%, and to rescale scores so that operating points transfer cleanly across file types. The model accounts for some of our empirical observations and provides concrete guidance for policy tuning and for benchmarking richer features or learning-based systems.

\textbf{(3) Empirical evaluation of CNN detectors across diverse ransomware configurations.} We systematically evaluate CNN-based detection methods by applying realistic intermittent encryption strategies derived from a diverse set of state-of-the-art ransomware families.%, including (but not limited to) BlackCat, Akira, and LockBit. 
%By considering multiple configuration options from these malware families, we comprehensively capture the varied ways ransomware impacts file systems at the byte-level. 

\section{Related research} 
Below, we discuss related research categorized into two segments: (1) File‑Content Classification Approaches, which examine how encryption reshapes a file’s byte‑level structure and use those changes to determine whether a file has been encrypted, and, (2) System‑ and Process‑Level Behavioural Defences, which detect or mitigate ransomware by monitoring program activity, I/O patterns, and other runtime signals rather than inspecting file contents directly.
\subsection{Encryption‑Induced File Structure and Content‑Based Classification}
%Encryption fundamentally alters a file’s byte-level structure, effectively randomizing its content. Cryptographic ciphers are designed so that ciphertext appears statistically indistinguishable from random data. Consequently, the distribution of byte values in a fully encrypted file tends to closely approximate a uniform distribution \cite{DeGaspari2022ReliableDetection}. This means an encrypted file exhibits near-maximal Shannon entropy (approaching 8 bits of entropy per byte), whereas ordinary unencrypted files (especially plain-text or other structured data) have lower entropy and more predictable byte patterns \cite{e24101503}. In practical terms, high entropy has long been recognized as a hallmark of encrypted content. For example, Davies et al. \cite{e24101503} note that “the closer a file’s overall entropy value is to eight bits, the higher the confidence that its contents are encrypted”. 

Past research has leveraged the properties of encrypted files to detect or characterize them. A common approach is to compute the file’s Shannon entropy as an indicator of randomness. If a file’s entropy is very high (near 8 bits/byte), it likely contains encrypted or compressed data \cite{DeGaspari2022ReliableDetection, e24101503}. Similarly, statistical tests have been used to compare a file’s byte-frequency distribution to the uniform distribution expected of ciphertext. For instance, chi-square tests on the byte-value distribution can help distinguish encrypted data from certain other data types \cite{DeGaspari2022ReliableDetection}. Other metrics like Kullback–Leibler divergence and serial byte correlation (which measures how predictable one byte is from the previous) have also been explored to identify encryption by the absence of structure in the byte stream. 

However, a challenge in practice is that encryption is not unique in producing high-entropy, random-looking files. Compression can similarly remove structural patterns and yield near-uniform byte distributions. For example, Microsoft's DOCX files, which are essentially ZIP archives containing a collection of XML files and other resources, often have randomness patterns mirroring those of encrypted data\cite{de2020encod}. This overlap leads to false positives when using measures such as entropy alone for detecting encryption. Recent studies have addressed this by employing more nuanced analyses or machine learning to differentiate truly encrypted content from merely compressed content based on subtle differences in their byte-level statistics \cite{Beebe}. 

For example, Kozachok and Spirin \cite{kozachok2021model} introduced a detection algorithm based on a statistical model that integrates multiple entropy and randomness tests. Their method demonstrated high effectiveness, achieving detection accuracies of up to 97\% in identifying encrypted content within files. 

Kim et al. \cite{kim2022byte} introduced lightweight indicators EntropySA and  DistSA which were analysis-based on a small sample area of the file. Blocks such as the header or footer are typically well-structured across files compared to other regions in the file; therefore, by examining this chunk, it is easier to realize distortion within its content by comparing it to the rest of the file. This method performed so well with over 99.5\% accuracy in distinguishing unencrypted vs fully encrypted files. This approach reduces the potential of false alarms and also cuts on computational overheads.

Casino et al.\cite{casino2019hedge} also extracted a rich set of features from file blocks and used a Random Forest classifier to differentiate encrypted vs non-encrypted. They leveraged features that examine more than just frequency to look at contextual data to distinguish from compressed data apart from cipher text.

\textbf{Limitations and how we differ.} Unlike prior work, we frame partial encryption as \textbf{a continuous}, file‑type–specific process and \textbf{study its full trajectory}. The vast majority of earlier works  focus only on the two endpoints: clear‑text and fully encrypted. For this reason they lack ground truth for the “blind zone’’ where intermittent ransomware hides. We close that gap by measuring how a series of methodically chosen file structure properties drift as encryption coverage increases from 0\% to 100\% across eleven common formats, producing the first empirical atlas of byte‑level erosion. We additionally provide an analytic framework that captures how a file’s statistical signature degrades as encryption advances. %The observed traits do not only have applications in the detection of file-encrypting ransomware, but should also serve broader applications including incident response triage, forensic analysis, and data ex-filtration detection among others. 

%To augment today’s largely empirical treatment of intermittent encryption, we introduce an analytic framework that captures how a file’s statistical signature degrades as encryption advances. The model produces closed‑form, file‑type–specific detectability ceilings and pinpoints regimes where entropy‑only detectors are provably inadequate—offering a theoretical lens that complements existing empirical findings. Finally, we operationalise these insights with chunk‑level CNNs trained on real intermittent‑encryption modes (as implemented in popular ransomware families such as, BlackCat, Akira, LockBit) and show they consistently outperform whole‑file models, establishing a robust baseline for future defences.

\subsection{System‑ and Process‑Level Behavioural Defences}
Complementary to file–content inspection, a large body of work seeks to identify ransomware \emph{while it runs}, by observing the program’s interactions with the operating system. 
Early approaches focused on sequences of file modification calls and signs of on-the-fly encryption. For example, UNVEIL \cite{kharaz2016unveil} watch for bursts of file write or rename operations combined with a sudden increase in data entropy of the written content. In these systems, each suspicious file operation (e.g. writing high-entropy data to a file) raises an anomaly score; once the score crosses a threshold (indicating enough damage done), the ransomware process is terminated. 

Later systems move this monitoring into the OS kernel for finer control and data recovery. Notably, ShieldFS \cite{continella2016shieldfs} is a ransomware-aware filesystem driver that intercepts low-level I/O requests (IRPs) and keeps copy-on-write backups of files in real time. It continuously learns normal application access patterns, and if a process’s file-access behavior deviates strongly (e.g. many small writes across numerous files in a short time), ShieldFS flags it as ransomware. The malicious process’s file operations can then be transparently rolled back using the saved copies. 

% A similar idea is RansomBlock \cite{ransomblock2021}, which surgically blocks suspicious file modifications; like ShieldFS, it operates in the kernel to intercept file-system calls and can restore original files once ransomware activity is confirmed.
%%%
% \cite{ransomblock2021}
% \cite{kharraz2017} 
%%%%%%%
% Another line of defense targets the encryption APIs themselves. Kharraz\emph{etal.}\cite{kharraz2017} observed that most Windows cryptoransomware samples invoke the OS’s built-in CryptoAPI functions (for key generation and encryption) right before writing encrypted data to disk. By hooking these cryptographic library calls, a security tool can detect and block ransomware in a deterministic way (since benign processes rarely encrypt hundreds of user files at once). This strategy, exemplified by tools like PayBreak, can reliably stop encryption by intercepting calls to CryptoAPI and even escrow the encryption keys for recovery. 

Finally, honeypot object plants are another family of defenses: files, registry entries, or network shares that should never be altered during normal operations. The idea is that any modification to these canaries is a clear sign of ransomware. The honey files layer in RansomWall \cite {shaukat2018ransomwall}, and Windows Defender’s Controlled Folder Access fall in this category.

% Finally, hypervisor‑level introspection proposes to inspect guest memory pages for high‑entropy buffers being flushed to disk\cite{vinokurov2022virt}; while promising for cloud workloads, such schemes demand specialised hardware support and are currently impractical for commodity endpoints.

\textbf{Limitations and how we differ.} Behavioural defences react \emph{during} encryption and therefore (i) struggle against \emph{slow‑burn intermittent ransomware} that throttles I/O below anomaly thresholds, (ii) require privileged hooks or kernel drivers that raise deployment barriers, and (iii) often detect only \emph{after} partial data loss. %\textbf{Our work addresses the complementary \emph{content} axis}: we show that carefully chosen byte‑histogram features and chunk‑level CNNs can expose intermittent encryption even when process‑level signals remain benign, offering a defence layer that is orthogonal to—and can be combined with—behavioural approaches.

Our work addresses the complementary \emph{content} axis: we inspect the post-attack artefact (the modified file itself), so our signal persists regardless of how the encryption was orchestrated (throttling/sleep, API unhooking, LOLBins, VM evasion). %This makes content inspection deployable without kernel hooks and suitable for retrospective scans on endpoints, servers, or backups, while behavioural defences remain better for stopping encryption early; together they form a defence-in-depth stack.

%%%%%%%%%%%%%%%%%%%%%%%%%%%

\section{Empirical landscape of partial‑encryption artefacts}
% This section presents our analysis of how intermittent encryption reshapes the byte-level structure of everyday file formats. We begin by describing the datasets used in our experiments, followed by details on how we replicate partial-encryption strategies observed in real-world ransomware families. Next, we summarize the statistical metrics used to track structural changes, and conclude with key findings that capture the full trajectory of file evolution as encryption coverage increases across diverse file types.

\subsection{Dataset Description}
\label{datas1}
% Our experiments are based on files sourced from NapierOne \cite{napierone}, an open and publicly available dataset. NapierOne is a mixed-file cybersecurity dataset developed primarily for ransomware detection and forensic analysis research. The dataset was created to address challenges in reproducibility and to enhance consistency by enabling research replication and repeatability. The dataset incorporates a variety of file types that collectively reflect the typical contents of personal and professional storage drives. 

Our experiments are based on files sourced from NapierOne \cite{napierone}, an open and publicly available dataset. Our selected dataset contains 26,482 files spanning a diverse set of file types grouped into three broad compression classes, capturing lossless and uncompressed formats, container-based documents that blend compressed and structured parts, and lossy media. 

\textbf{(1) Lossless or Uncompressed formats} include BMP, PNG, and TXT files (6,400 samples), which provide predictable headers and structure, high post-compression entropy, or strong byte-level regularity.

(\textbf{2) Container (Variable Compression) formats} consist of DOC, DOCX, PDF, PPT, and XLS files (11,330 samples), each encapsulating multiple streams such as text, XML parts, embedded media, and compressed subcomponents resulting in heterogeneous internal compression behaviors. 

(\textbf{3) Lossy formats} include JPG images, MP3 audio, and MP4 video files (8,752 samples), all dominated by transform-based entropy coding that produces near-uniform byte distributions. 

% In total, our dataset contains 26,482 files spanning a wide variety of structures, compression behaviors, and media types. This diversity captures a broad range of baseline entropy and byte regularity seen in real-world data: lossless and uncompressed files retain clear structure, lossy media often appear close to random at the byte level even without encryption, and container documents vary widely because a single file may combine uncompressed text, deflate-compressed XML, and already compressed images or audio. Covering all three groups allows us to analyze how partial encryption interacts with these native byte distributions.
%%%%%%%%%%%%%%%%%%%%%%%%%%%%%%%%%%%%%%%%

\begin{table}[h!]
\scriptsize
\centering
\caption{Modes of Intermittent Encryption Used by Ransomware Families}
\begin{tabular}{lcccc}
\toprule
 
\diagbox[width=16em]{\textbf{Ransomware Family}}{\textbf{Mode}}
    & \textbf{Adaptive}
    & \makecell{\textbf{Dot}\\\textbf{Pattern}}
    & \textbf{Head}
    & \textbf{Hybrid} \\
\midrule
Agenda (Qilin) (2022) \cite{Toulas_2022}       &             & \checkmark & \checkmark & \checkmark \\
BlackCat (ALPHV) (2021) \cite{cyberark2025}    & \checkmark  & \checkmark & \checkmark & \checkmark \\
Qyick (2022) \cite{Toulas_2022}                & \checkmark  &            &            &             \\
Play (2022) \cite{Milenkoski_Walter_2025}      & \checkmark  &            &            &             \\
Royal (Black Suit) (2022) \cite{Langley_2025}  &             & \checkmark &            &             \\

Akira (2023) \cite{CISA2024Akira}               & \checkmark  & \checkmark & \checkmark & \checkmark \\

DarkSide (2020) \cite{Loman_2021}              &             &            & \checkmark &             \\
Black Matter (2021) \cite{Loman_2021}          &             &            & \checkmark &             \\
LockBit 2.0/3.0/4.0 (2021) \cite{CISA2023LockBit}  &             &            & \checkmark &             \\
\bottomrule
\end{tabular}
\label{encryption-schemes-used}
\end{table}

%%%%%%%%%%%%%%%%%%%%%%%%%%%%%%%%%%%%%%%
\subsection{Implementing Intermittent Encryption}
Our encryption design closely mirrors techniques observed in modern file-encrypting ransomware. We use the Advanced Encryption Standard (AES) in Galois/Counter Mode (GCM) with 128-bit keys, selected for its widespread use in cryptographic ransomware attacks \cite{aes-gcm}.

To evade detection, ransomware typically applies encryption in a non-uniform manner, selectively targeting portions of files based on predefined logic. These intermittent strategies vary in terms of which parts of the file are encrypted and how the encryption is distributed across the file. We reviewed a range of recent ransomware families and replicated several commonly observed schemes.
Table~\ref{encryption-schemes-used} summarizes the encryption patterns we implemented and the ransomware families associated with each approach. A brief description of each scheme is provided below.

\textbf{(1) Head}: A simple strategy where only the initial $X$ bytes (or “head”) of the file is encrypted, \textbf{(2) Dot Pattern}: A periodic encryption pattern, that breaks the file into $N$ fixed size blocks, and for each block, encrypts 
$X$ bytes, resulting in a “dotted” appearance in the encrypted data.  \textbf{(3) Hybrid}: A composite approach that combines multiple strategies (e.g., head + dot pattern), \textbf{(4) Adaptive}: A dynamic strategy where the ransomware adjusts choice of the above modes, and even full encryption, based on factors such as file size, file type, or system performance. Details of parameters used in our experiments will be described where each mode is used. %For instance, large files may be only partially encrypted to reduce runtime while still achieving disruption.

\begin{figure}
    \centering

    \begin{subfigure}[b]{0.45\columnwidth}
        \centering
        \includegraphics[width=\linewidth]{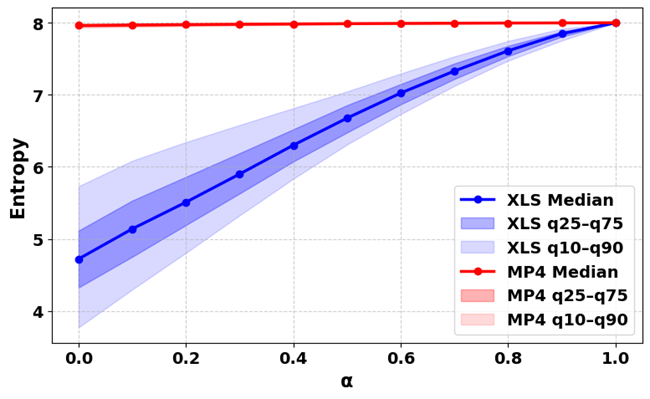}
        \caption{Entropy}
        \label{entropy}
    \end{subfigure}
    \hfill
    \begin{subfigure}[b]{0.45\columnwidth}
        \centering
        \includegraphics[width=\linewidth]{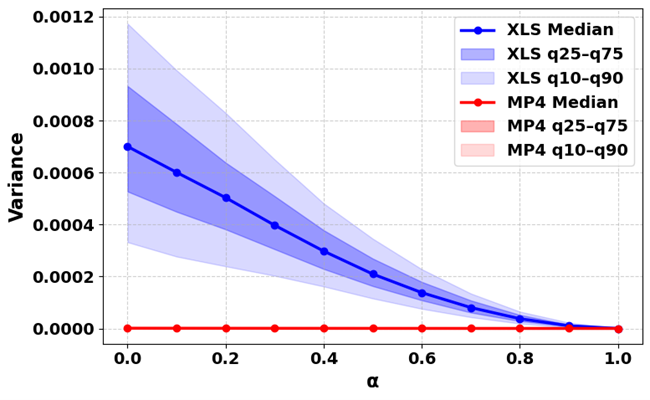}
        \caption{Variance}
        \label{kl}
    \end{subfigure}

    \vspace{0.2em}

    \begin{subfigure}[b]{0.45\columnwidth}
        \centering
        \includegraphics[width=\linewidth]{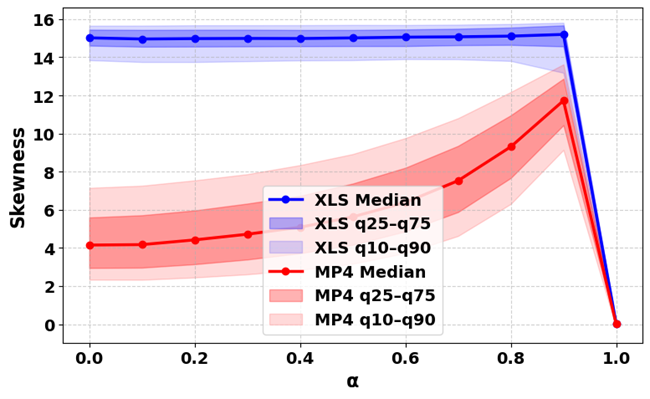}
        \caption{Skewness}
        \label{skewness}
    \end{subfigure}
    \caption{Evolution of statistical traits of two representative file families (xls which is uncompressed/binary, and mp4 which is lossy) with increasing amounts of encryption. The two inner bands around each curve represent the first and third quartiles, while the external bands represent the tenth and ninetieth percentiles.}
    \label{fig:stats}
\end{figure}
% \begin{figure*}[ht]
%     \centering
%     \begin{subfigure}[b]{0.32\textwidth}
%         \centering
%         \includegraphics[width=\textwidth]{Figures/entropy.png}
%         \caption{Entropy}
%         \label{entropy}
%     \end{subfigure}
%     \hfill
%     \begin{subfigure}[b]{0.32\textwidth}
%         \centering
%         \includegraphics[width=\textwidth]{Figures/variance.png}
%         \caption{Variance}
%         \label{kl}
%     \end{subfigure}
%     \hfill
%     \begin{subfigure}[b]{0.32\textwidth}
%         \centering
%         \includegraphics[width=\textwidth]{Figures/skewness.png}
%         \caption{Skewness}
%         \label{skewness}
%     \end{subfigure}
%     \caption{Evolution of statistical traits of two representative file families (xls which is uncompressed/binary, and mp4 which is lossy) with increasing amounts of encryption. The two inner bands around each curve represent the first and third quartiles, while the external bands represent the tenth and ninetieth percentiles.}
%     \label{fig:stats}
% \end{figure*}

\begin{table*}[ht]
\scriptsize
\centering
\caption{Sen’s slope (per 0.1 increase in $\alpha$) for the median and Inter-quartile Range (IQR) series of each metric, with one-sided Mann–Kendall \emph{p}-values (alternative hypothesis indicated by arrows: $\uparrow$ increasing, $\downarrow$ decreasing). Brackets show 95\% confidence intervals.}
\label{tab:senslope_mk}
\begin{tabular}{llcccc}
\toprule
\multirow{2}{*}{\textbf{Filetype}} & \multirow{2}{*}{\textbf{Metric}} & \multicolumn{2}{c}{\textbf{Median series}} & \multicolumn{2}{c}{\textbf{IQR series}} \\
\cmidrule(lr){3-4} \cmidrule(lr){5-6}
 &  & \textbf{Sen's slope [95\% CI]} & \textbf{MK $p$-value} & \textbf{Sen's slope [95\% CI]} & \textbf{MK $p$-value} \\
\midrule
MP4 & Entropy  & 0.0036 [0.0029, 0.0044]      & 1.31e-05 ↑ & -0.0025 [-0.0031, -0.0021] & 1.31e-05 ↓ \\
MP4 & Skewness & 0.4914 [0.2482, 0.9174]      & 0.004 ↑    & 0.0784 [-0.1840, 0.1179]    & 0.175 ↑ \\
MP4 & Variance & -9.31e-08 [-1.14e-07, -7.29e-08] & 1.31e-05 ↓ & -7.51e-08 [-9.30e-08, -5.73e-08] & 1.31e-05 ↓ \\
XLS & Entropy  & 0.3488 [0.3096, 0.3790]      & 1.31e-05 ↑ & -0.0821 [-0.0908, -0.0752]  & 1.31e-05 ↓ \\
XLS & Skewness & 0.0188 [-0.00551, 0.02991]   & 0.043 ↑    & 0.00141 [-0.01035, 0.01853] & 0.500 ↑ \\
XLS & Variance & -7.67e-05 [-9.37e-05, -5.74e-05] & 1.31e-05 ↓ & -3.94e-05 [-5.14e-05, -2.82e-05] & 1.31e-05 ↓ \\
\bottomrule
\end{tabular}
\end{table*}

\subsection{Statistical Traits Studied}
To capture how intermittent encryption reshapes file structure, we evaluate three complementary statistical measures at each encryption fraction ($\alpha$). These metrics, drawn from information theory and descriptive statistics, jointly provide a detailed view of the transition from plaintext to full ciphertext. For robustness, we accompany our observations with statistical significance testing.

The three measures are Shannon entropy, variance, and skewness, each offering a distinct yet related perspective.

\textit{Shannon entropy} quantifies the unpredictability of byte sequences. In theory, entropy should rise as encryption coverage increases, approaching the maximum for ideal ciphertext. Our analysis focuses on the rate of this rise across different file types.

\textit{Variance} captures how unevenly byte values are distributed. Structured plaintext typically shows high variance due to characteristic byte-frequency peaks, whereas ciphertext exhibits minimal variance under a uniform distribution. We examine how quickly variance declines with increasing encryption coverage.

\textit{Skewness} measures asymmetry in byte-frequency distributions. Many plaintext formats exhibit marked skew, reflecting domain-specific symbol usage, while encryption aims to remove this bias. We study how skewness trends toward zero as encryption progresses.

% While our broader study of intermittent encryption includes all strategies described in Section~\ref{datas1}, the statistical analyses in this section focus on the head-pattern scheme for simplicity. For the metrics considered here, only the proportion of encrypted bytes influences the results; the spatial or temporal arrangement of encrypted versus unencrypted regions has no effect.

Together, these three metrics provide complementary views of structural change, each highlighting a different aspect of how byte distributions evolve. To formalize the patterns, we use the Mann–Kendall~\cite{kendall1975rank} test and Sen’s slope~\cite{sen1968}. Mann–Kendall is a nonparametric, rank-based test for a monotonic trend that makes no normality or linearity assumptions; we use one-sided alternatives consistent with expectations (e.g., increasing for entropy, decreasing for variance) and report the resulting \emph{p}-values. Sen’s slope complements this by estimating the median pairwise slope, providing a robust effect size in interpretable units with confidence intervals. We apply both procedures to the median series to characterize trends in central tendency and dispersion.

\subsection{Empirical Observations on Evolution Trajectories of Files due to Encryption}
\paragraph{Observations.}
Figure \ref{fig:stats} and Table \ref{tab:senslope_mk} illustrate the observed statistical traits using two \emph{representative} file families that bracket the spectrum: \textbf{XLS} (highly structured, lossless) and \textbf{MP4} (already near-uniform, lossy). 
% Full tables for all file types appear in the Appendix (see Section \ref{append}). 

For \emph{entropy}, the median curves rise with encryption coverage for both families, and one-sided Mann–Kendall tests under an increasing alternative hypothesis confirm monotonicity (significant in both cases). While these general trends are not surprising, the crucial insight is in the \emph{rate}. The effect sizes differ by orders of magnitude: Sen’s slope is approximately \(+0.35\) bits per \(0.1\) increase in \(\alpha\) for XLS (95\% CI \([0.3096,\,0.3790]\)) but only \(+0.0036\) for MP4 (95\% CI \([0.0029,\,0.0044]\)). Visually, MP4’s entropy curve is nearly flat; however the test detects a small but consistent upward drift that is statistically significant. 

\emph{Variance} shows the mirror image of entropy’s behavior: under a decreasing alternative hypothesis, there is a clear decline for XLS (about \(-7.67\times10^{-5}\) per \(0.1\); 95\% CI \([{-}9.37\times10^{-5},\,{-}5.74\times10^{-5}]\)) and a near-zero change for MP4 (\(\approx{-}9.31\times10^{-8}\) per \(0.1\)). The shaded bands echo this asymmetry: the entropy IQR contracts steadily for XLS (about \(-0.082\) per \(0.1\); 95\% CI \([{-}0.0908,\,{-}0.0752]\)), indicating convergence toward uniform byte usage, while MP4’s IQR shrinks only slightly (\(\approx{-}0.0026\) per \(0.1\)). 

\emph{Skewness} behaves as a secondary, format-dependent cue: MP4 exhibits a modest increasing trend (significant one-sided test; Sen’s slope \(\approx +0.49\) per \(0.1\), 95\% CI \([0.248,\,0.917]\)), whereas XLS shows little systematic movement (CI spans zero). A final detail is the abrupt drop in skewness between $\alpha=0.9$ and $1.0$. Skewness tracks asymmetry in the byte histogram: as long as a thin residue of plaintext remains, a handful of byte values dominate and the statistic stays elevated. Once coverage reaches 100\%, modern ciphers yield an effectively uniform histogram and skewness collapses toward zero. 

Notably, the \emph{shape} of this drop differs by family: in \textbf{XLS}, skewness remains high and essentially flat until very late and then falls off a cliff, indicating that strong symbol biases persist until the last pockets of plaintext are removed; in \textbf{MP4}, skewness first drifts upward and only then collapses. This contrast reinforces that skewness carries file-type–specific signals that should be useful late in the trajectory for structured formats, but less reliable as a standalone cue for near-uniform highly compressed files.
Finally, because we use a head-only scheme and these aggregate histogram statistics depend primarily on the \emph{fraction} encrypted rather than the spatial layout, the traits seen can be read as coverage-driven, family-specific curves. 

Comprehensive statistics for all file families appear in the Appendix (Section \ref{append}). For completeness, we additionally report three distance-to-plaintext measures (Euclidean ($L2$) distance, Kullback–Leibler ($KL$) divergence, and total variation ( $TV$)) in the Appendix; they track the same monotone progression and mainly serve as sanity checks, with $KL$ linking to the theory developed in the next section.

% These statistics are complemented with three \textit{distance-to-plaintext measures}: Euclidean (L2) distance, Kullback–Leibler (KL) divergence, and total variation (TV). Each measures how far a partially encrypted file's byte distribution deviates from its original cleartext state.

\paragraph{Implications of Observed Traits.}
First, global histogram cues (entropy and variance) offer substantial leverage on structured families like XLS but far less on near-uniform media like MP4; thresholds tuned on one family will not transfer cleanly to another. Second, the steady band narrowing for structured types suggests that decision confidence improves with coverage, while the minimal band change in near-uniform types limits the utility of whole-file statistics. Practically, this motivates file-type–aware thresholds and complementary localized analysis (for example, chunk-level CNNs) to capture weak, spatially concentrated evidence that global histograms miss.

These regularities motivate a compact, first-principles description that maps encryption coverage to observable byte histograms in a file-family–specific way and clarifies when histogram-based detection will saturate. In the next section, we formalize this by modeling partial encryption as a mixture and analyzing divergence from uniformity via KL, yielding family-dependent thresholds that explain the empirical curves and inform principled, file-type–aware decision rules.

\section{Modeling File Dynamics of Intermittent Encryption}
\label{sec:stat-model}
Guided by the observations in the previous section, we now develop a simple analytic model for partial encryption that operates directly on byte histograms. %A partly encrypted file is treated as a convex mixture of ciphertext and its original distribution, parameterized by the encrypted fraction. This model explains the shapes seen empirically and provides explicit limits for histogram based detection.

\subsection{Formulating encrypted file system as a mixture density}
\label{formulate}
Intermittent‑encryption ransomware encrypts only a portion of each file,
leaving the remainder in cleartext.  
The observable byte‑value distribution is therefore a blend of two
sources:

(i) Ciphertext bytes. 
        For a strong cipher such as AES, ciphertext is empirically
        uniform: every byte value \(b\in\{0,\dots,255\}\) occurs with
        probability \(1/256\).  
        This distribution is given by,
        \(
          P_{\text{enc}}(b)=U(b)=\frac{1}{256}.
        \)

          (ii) The untouched part of the file retains its pre‑attack
        distribution, denoted \(P_{\text{orig}}(b)\).

If the malware encrypts a fraction, \(\alpha\)
(\(0\le\alpha\le1\)) of bytes, the resulting data distribution can be modeled as a convex mixture with density function,

\begin{equation}
  P_{\text{mix}}(b;\alpha)
    \;=\;
    \frac{\alpha}{256}
    \;+\;
    (1-\alpha)\,P_{\text{orig}}(b).
  \label{eq:pmix}
\end{equation}

When \(\alpha = 1\), the file is fully encrypted, meaning its byte histogram
        is perfectly uniform. When \(\alpha = 0\), no encryption has occurred and so
        the histogram equals \(P_{\text{orig}}\). From a defender’s standpoint, the case where \(\alpha = 1\) causes the most suspicion since the final uniform distribution, \(P_{\text{enc}}\), is a solid signature of file encryption. 
 
An attacker, therefore, tries to \emph{lower \(\alpha\)} just enough that
\(P_{\text{mix}}(b; \alpha)\) drifts sufficiently far from uniform to
evade entropy‑based or uniformity‑based detectors, while still encrypting
enough data to significantly corrupt the file.

\subsection{Escape trajectory and a family‑dependent detectability ceiling}
 Central to our model, we introduce the notion of the \emph{escape trajectory}: the
        curve
        describing how quickly \(P_{\text{mix}}(\alpha)\) diverges from
        the uniform reference \(P_{\text{enc}}\) as \(\alpha\) decreases
 from 1.  This trajectory captures how much encryption an attacker
        must relinquish before the file stops resembling ciphertext, potentially evading detection. Below, we discuss our reasoning behind the choice of trajectory function, before leveraging it to formalize bounds for family-detectability ceiling, and discussing its implications to classifier design. \\
\subsubsection{Defining the file escape trajectory}
\label{Robs}

%Our focus here is to characterize the attacker’s room for manoeuvre. 
        %We study how this rate changes across two broad and representative file‑type
        %families: near‑uniform lossy files (e.g., jpeg images and mp3 files), and highly structured, uncompressed files (e.g., text/source files). 

        Let
\(D:\mathcal{P}\times\mathcal{P}\to[0,\infty)\)
be any non-negative \emph{dissimilarity measure} that equals
\(0\) only when its two arguments are identical.
We impose no further conditions (such as symmetry or the triangle
inequality), so $D$ may be a true metric or an asymmetric
divergence.

        We define the \emph{escape trajectory} as the mapping
        \(
          \mathcal{E}(\alpha)
          \;=\;
          D\bigl(P_{\text{mix}}(b;\alpha)\,\|\,P_{\text{enc}}(b)\bigr),
          \qquad \alpha\in[0,1].
        \)
        It starts at \(\mathcal{E}(1)=0\) and rises as \(\alpha\)
        decreases from 1, describing how quickly
        \(P_{\text{mix}}(b;\alpha)\) diverges from the uniform reference.  
        
        For the measure \(D\), we use the Kullback--Leibler (KL) divergence (or relative entropy) due to its direct entropy interpretation. In this setting KL is literally the remaining
``entropy shortfall'', i.e., the number of bits of randomness per byte that the block still
has not gained because some plaintext structure is leaking through. As intermittent
encryption progresses, that shortfall shrinks; KL tracks exactly how much structured
(non-random) signal is left. A plain geometric distance like $\ell_{2}$ can tell you
``different'' but not ``how many bits of randomness are still missing,'' so it lacks
the immediate interpretability we need for leak reasoning and threshold setting.

\begin{table}[ht]
\scriptsize
\centering
\caption{Family constants $c_F^{2}$ estimated from cleartext files. For each file type or family, we repeatedly draw a random subset of 200 files (100 replicates), compute $c_F^{2}$ for each replicate, and report the median with a 95\% confidence interval across replicates.}
\label{tf-values}
\begin{tabular}{l l r l}
\toprule
\textbf{Compression Type} & \textbf{Filetype} & Median \textbf{$c_F^{2}$} & \textbf{95\% CI} \\
\midrule
\multirow{5}{*}{Uncompressed}
    & bmp & 0.005879 & [0.004812, 0.007345] \\
    & txt & 0.058256 & [0.057612, 0.058641] \\
    & xls & 0.179274 & [0.172506, 0.189904] \\
    & doc & 0.073060 & [0.062462, 0.086819] \\
    & ppt & 0.006465 & [0.005152, 0.008660] \\
\midrule
\multirow{6}{*}{Compressed}
    & docx & 0.001245 & [0.000975, 0.001596] \\
    & mp3  & 0.000222 & [0.000204, 0.000248] \\
    & jpg  & 0.000129 & [0.000122, 0.000136] \\
    & mp4  & 0.000250 & [0.000232, 0.000271] \\
    & pdf  & 0.000733 & [0.000647, 0.000945] \\
    & png  & 0.000082 & [0.000075, 0.000089] \\
\bottomrule
\end{tabular}
\end{table}

\subsubsection{Per-Family Detection Ceilings}
We present a proposition that specialises the textbook KL–to–$\chi^{2}$ bound
to our ransomware mixture
model and per-family leak dynamics.  While the bound itself is standard; our novelty is the specialization to ransomware’s convex mixture with uniform ciphertext, the associated estimation of file family-specific constants 
from cleartext corpora, and the policy-level uses (family-aware thresholds, score normalization and file chunking). 
%A corollary then lifts the same bound to a full host whose byte
%histogram aggregates multiple file families.

%Leveraging the standard KL–to–$\chi^2$ upper bound \cite{CoverThomas2006,DragomirGluscevic2000}, we obtain the
%following information‑theoretic ceiling on any KL‑based detector.

%%237795670
\begin{proposition}[Family–specific unattainable threshold]
\label{thm:family-threshold}
Fix a file type family, $F$, with original byte histogram \(P_{\text{orig},F}\) and define
\begin{align}
\Delta(b)            &= P_{\text{orig},F}(b)-\tfrac{1}{256}, \\
\varepsilon_b(\alpha)&= 256(1-\alpha)\,\Delta(b), \\
c_F^{2}\;&=\;\bigl\|P_{\text{orig},F}-U\bigr\|_{2}^{2}.
\end{align}
Assume the high–encryption (small‑leak) condition
$\max_{b}|\varepsilon_b(\alpha)|<1$ holds.
If a KL–based detector for this family triggers only when its score exceeds
a threshold $\tau_{\!F}$, then any choice
\(
\tau_{\!F} \;>\; \frac{256}{\ln 2}\,c_F^{2}\,(1-\alpha)^{2}
\)
is information–theoretically guaranteed not to detect the attack. No partially
encrypted block (with admissible $\alpha$) can ever cross the threshold.
\end{proposition}

Below, we present a sketch of the proof and leave the full proof to the Appendix (Section \ref{append}), Page \pageref{append}.

\begin{proof}
%The proof is basically an instantiation of the KL envelope \cite{CoverThomas2006,DragomirGluscevic2000} in the context of our ransomware scenario. 
Starting from
the exact KL expression for $P_{\text{mix}}(\alpha)$, we factor out the
uniform baseline, introduce $\varepsilon_b(\alpha)=256(1-\alpha)\Delta(b)$,
and use the zero–mean property $\sum_b\Delta(b)=0$ to cancel the linear
term.  Applying the inequality
$\log_{2}(1+x)\le x/\ln 2$ for $|x|<1$ coordinate‑wise (guaranteed by
$\max_b|\varepsilon_b(\alpha)|<1$) yields an upper bound proportional to
$\sum_b \varepsilon_b(\alpha)^2$.  Substituting
$\varepsilon_b(\alpha)=256(1-\alpha)\Delta(b)$ and collecting constants
gives the ceiling $\tfrac{256}{\ln 2}c_F^{2}(1-\alpha)^2$.

\end{proof}

%\paragraph{Interpretation, scope, and contribution.}
Proposition~\ref{thm:family-threshold} leverages a classical KL to chi-square envelope \cite{DragomirGluscevic2000} and tailors it to intermittent encryption modeled as a convex mixture with a uniform ciphertext reference. The small-leak condition \(\max_b|\varepsilon_b(\alpha)|<1\) holds in the high-encryption regime that attackers tend to prefer, since they aim to encrypt most bytes while leaving only a thin residue of plaintext to evade detectors; for highly compressed families where \(\Delta(b)\) is small, the condition persists over a wider range of \(\alpha\). The ceiling is expressed through a measurable, family-specific constant \(c_F^2=\|P_{\text{orig},F}-U\|_2^2\), which links the inequality directly to data and to the empirical trajectories reported earlier. Our core contribution lies in specializing the envelope to the ransomware mixture setting, reducing it to the single statistic \(c_F^2\) that can be estimated from cleartext, and translating that quantity into per-family thresholds and \(\alpha\) regimes developed in the next subsection.
\begin{figure}[!t]
    \centering
    \includegraphics[width=0.60\linewidth]{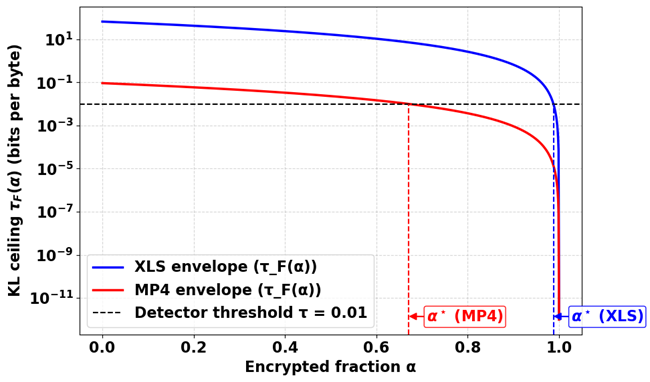}
    \caption{Per-family escape ceilings under intermittent encryption.
  Theoretical KL ceiling $\tau_F(\alpha)=\tfrac{256}{\ln 2}\,c_F^{2}(1-\alpha)^2$
  for XLS (blue) and MP4 (red); dashed line marks detector threshold $\tau=0.01$.}
    \label{envelope}
\end{figure}

\subsection{Implications for KL-based detectors}

The bound in Proposition~\ref{thm:family-threshold} yields a simple operating rule: for any detector threshold $\tau$, the onset of guaranteed non-detection occurs at a family specific coverage $\alpha^\star$ given by
\(
\alpha^\star \;=\; 1 - \sqrt{\tfrac{\ln 2}{256}}\;\tfrac{\sqrt{\tau}}{c_F}.
\)
When $\alpha>\alpha^\star$, no histogram rule dominated by KL can trigger on that family. For example, XLS has \(c_F^{2}\!\approx\!1.79\times10^{-1}\), TXT about \(5.83\times10^{-2}\), whereas MP4 is near \(2.50\times10^{-4}\) and JPG near \(1.29\times10^{-4}\). These values differ by several orders of magnitude, so under the same \(\tau\) the implied \(\alpha^\star\) for XLS is much larger than for MP4 or JPG. 

Figure~\ref{envelope} visualizes this effect on a logarithmic \(y\)-axis for readability: the solid curve shows the threshold relation and the dotted vertical markers indicate \(\alpha^\star\) for XLS and MP4 under a common illustrative \(\tau\). The XLS marker lies far to the right, reflecting its higher \(c_F^{2}\) and longer detectability window, whereas the MP4 marker lies much farther left.

Two practical consequences follow. First, when operating with a fixed \(\tau\), consult the family’s \(c_F^{2}\): higher values support detection at higher coverages, lower values require looser thresholds or additional features beyond histograms. Second, to align operating points across file types, normalize the KL score by \((256/\ln 2)\,c_F^{2}\) and threshold the normalized statistic; this collapses family differences into the coverage term and simplifies cross-family policy tuning.

\section{Building Classification Engine}
\subsection{Overview}
Beyond advancing the general understanding of how intermittent encryption interacts with everyday files on user devices, our empirical atlas and analytic model reveal actionable design directions that guide how we build the classification engine.
The empirical study shows that (i) global byte-histogram cues are strongly file-type dependent, (ii) their utility diminishes on near-uniform media as encryption coverage grows, and (iii) residual evidence becomes increasingly \emph{local}; so thresholds tuned on one family do not transfer cleanly to others. Complementing this, the KL-based analytic model offers a compact information-theoretic perspective: it quantifies entropy shortfall (bits per byte) relative to the uniform 256-bin ciphertext distribution, and induces per-family detectability ceilings. 

While our classification algorithm of choice (a Convolutional Neural Network (CNN)) learns its decision thresholds automatically, these insights still inform the architecture and training regimen: include diverse file families, cover multiple intermittent-encryption patterns, and emphasize locality where global statistics weaken. Guided by this, our classification study proceeds in stages: (i) a file-level CNN baseline, (ii) the same model with GAN-based augmentation to probe robustness in high-coverage, low-signal regimes, and (iii) a chunk-level CNN to explicitly test the locality hypothesis suggested by our earlier results.

%\[h[b] = \frac{1}{N}\sum_{i=1}^{N} \mathbf{1}_{\{x_i = b\}}, \qquad b=0,\dots,255,\]
\subsection{Byte–Histogram Image Encoding}
\label{sec:encoding}
CNNs expect image tensors, while our corpus contains many non-image files (e.g., DOCX, XLS, MP3, MP4, and TXT). We therefore represent each file as a fixed $16{\times}16$ single-channel image derived from its byte-value histogram. For a byte sequence $x_1,\dots,x_N$, we count the occurrences $\eta[b]$ of each byte value $b\in\{0,\dots,255\}$, normalize the counts by the largest bin $\eta_{\max}=\max_b\eta[b]$, and compute $\nu[b]=255\,\eta[b]/\eta_{\max}$. The resulting 256-dimensional vector is then reshaped in row-major order into an image $H$, where the pixels correspond sequentially to byte values \texttt{0x00}--\texttt{0xFF}. For whole-file models, we compute one image $H$ per file, while for chunk-level models, we apply the same mapping to each fixed-size chunk, producing a sequence $\{H^{(k)}\}$ whose predictions are subsequently aggregated at the file/host level.

\begin{figure}
    \centering

    \begin{subfigure}[b]{0.45\columnwidth}
        \centering
        \includegraphics[width=\linewidth]{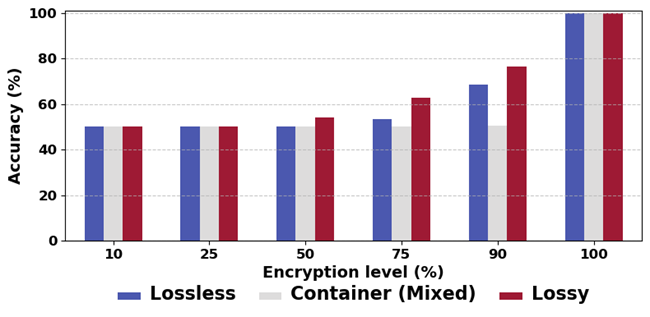}
        \caption{Baseline.}
        \label{baseline}
    \end{subfigure}
    \hfill
    \begin{subfigure}[b]{0.45\columnwidth}
        \centering
        \includegraphics[width=\linewidth]{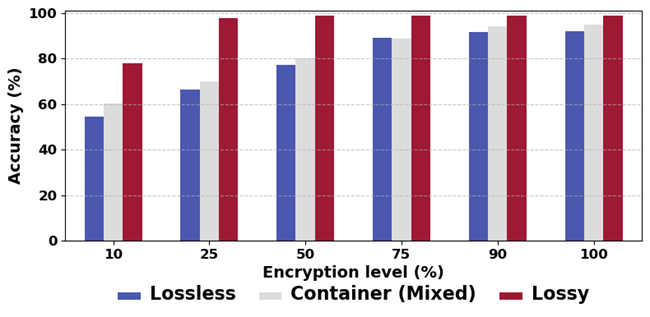}
        \caption{Baseline + intermittent.}
        \label{intermittent}
    \end{subfigure}

    \vspace{0.2em}

    \begin{subfigure}[b]{0.45\columnwidth}
        \centering
        \includegraphics[width=\linewidth]{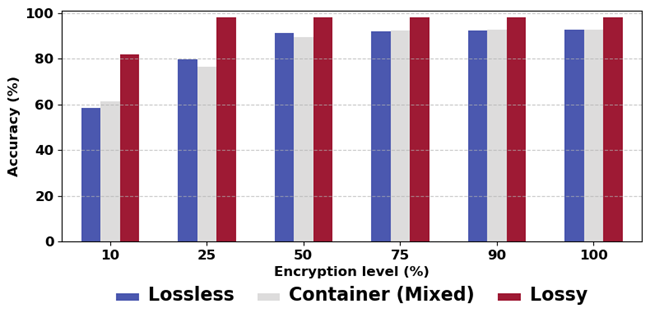}
        \caption{Baseline + GAN augmentation.}
        \label{gan-model}
    \end{subfigure}
    \caption{Classification accuracy for three detector-training strategies: (a) a baseline model trained only on fully encrypted and unencrypted files; (b) model additionally trained on real intermittently encrypted samples; and (c) baseline further augmented with GAN-synthesised intermittently encrypted examples.}
    \label{fig:base-intermittent-gan}
\end{figure}
% \begin{figure*}[t]
%     \centering
%     \begin{subfigure}{0.32\linewidth}
%         \centering
%         \includegraphics[width=\linewidth]{Figures/baseline.png}
%         \caption{Baseline: trained on fully encrypted and unencrypted.}
%         \label{baseline}
%     \end{subfigure}\hfill
%     \begin{subfigure}{0.32\linewidth}
%         \centering
%         \includegraphics[width=\linewidth]{Figures/Intermittent.png}
%         \caption{Baseline + real intermittently encrypted samples.}
%         \label{intermittent}
%     \end{subfigure}\hfill
%     \begin{subfigure}{0.32\linewidth}
%         \centering
%         \includegraphics[width=\linewidth]{Figures/gan.png}
%         \caption{Baseline + GAN-synthesised intermittent augmentation.}
%         \label{gan-model}
%     \end{subfigure}
%     \caption{Classification accuracy for three detector-training strategies: (a) a baseline model trained only on fully encrypted and unencrypted files; (b) model additionally trained on real intermittently encrypted samples; and (c) baseline further augmented with GAN-synthesised intermittently encrypted examples.}
%     \label{fig:base-intermittent-gan}
% \end{figure*}

\subsection{Details of Neural Networks used in our Pipeline}
\subsubsection{Architecture of Main CNN}

For the primary classification system, we screened several backbones (e.g., ResNet-18, MobileNetV3, EfficientNet-B0, ConvNeXt-Tiny) on a stratified validation split; \textbf{ConvNeXt-Tiny} \cite{liu2022convnet} delivered the best performance on our grayscale histogram images, so we use it for all reported results.

Each file/chunk yields a $16{\times}16$ single-channel histogram image (Sec.~\ref{sec:encoding}). To match the ConvNeXt stem, we upsample to $32{\times}32$ and replicate the channel to RGB; we then apply standard per-channel normalization. The network is trained \emph{from scratch} (ImageNet pretraining offered no benefit in our setting), and we replace the classifier head with a two-logit layer for binary classification. We train for 25 epochs with cross-entropy loss and AdamW (learning rate $1\!\times\!10^{-4}$). No geometric or photometric augmentations are used, as they would alter the semantics of histogram pixels.

\subsubsection{Architecture of GAN used for Augmentation}
\label{gan-arch}
In a subset of our experiments, we used a Generative Adversarial Network (GAN) to study the impact of data augmentation on system performance. A single GAN trained on 58,676 files produced unsatisfactory samples because of the distinct statistical characteristics of each file type. We therefore trained 11 file-type-specific GANs using a compact DCGAN-style architecture tailored to $16{\times}16$ single-channel images \cite{radford2015unsupervised}. Each model uses a 100-dimensional latent vector, LeakyReLU activations, and adversarial training with binary cross-entropy and Adam on mini-batches of size 64. Inputs are normalized to $[-1,1]$, while generator and discriminator learning rates are tuned separately for each file type.
\section{Performance Evaluation}
\subsection{Basic Detection Baseline}
To benchmark attacker advantage, we first trained a conventional content-based classifier \emph{not} designed for intermittent encryption: it learns from only the two endpoints (pre-encryption vs.\ fully encrypted) and is evaluated at intermediate coverages. On a balanced test set, this baseline separates 0\% vs.\ 100\% encryption perfectly (100\% accuracy across all file types).

Figure~\ref{baseline} shows what happens once an attacker encrypts only a fraction of each file. Accuracy degrades monotonically as coverage decreases, dropping from the fully encrypted case to roughly \(\sim\!50\%\) at 10\% coverage. Averaged over file types and coverages, the baseline attains \textbf{61.48\%} accuracy. 
The key takeaway is that a detector trained only on endpoints is brittle: intermittent encryption can halve accuracy while still causing substantial file damage, motivating models that explicitly learn mid-coverage regimes.

\subsection{Intermittent Encryption–Aware Baseline}

To address the brittleness of the endpoint-only baseline, we retrained the classifier with a diverse set of \emph{intermittent-encryption} examples spanning multiple coverage levels and patterns across all file types. Figure~\ref{intermittent} reports accuracy vs.\ coverage by file class. Performance lifts markedly across mid–high coverages; near the fully encrypted end it is comparable to the 100\% case. The macro-average accuracy over file types and coverages rises from 61.48\% to \textbf{84.36\%}.

Residual degradation remains at very low coverage (e.g., 10–25\%): the encrypted portion is too small relative to native file variability, so global histograms are dominated by plaintext and the signal becomes weak, especially for highly structured (lossless) formats. 

The takeaway here is that training on intermittent examples closes most of the gap, but low-coverage regimes still require features that capture \emph{local} evidence (addressed next with chunk-level models/augmentation).

\subsection{GAN-Augmented Intermittency-Aware Model}
We next augmented the intermittency-aware classifier with synthetic training examples generated by the per-type GANs (Section.~\ref{gan-arch}). Synthetic histogram images were mixed with real samples within each mini-batch; validation and test sets remained \emph{real-only}. 

Figure~\ref{gan-model} shows accuracy vs.\ coverage. The overall pattern mirrors the intermittency-aware baseline in Fig.~\ref{intermittent}: mid–high coverages are strong, while very low coverages remain challenging. Gains from GAN augmentation were small and inconsistent across file classes, and aggregate metrics were not significantly different from the real-only model. 
% In some cases we observed minor regressions at low coverage, consistent with synthetic artifacts not fully matching the variability of real files.

Overall, per-type GAN augmentation did not materially improve detection under intermittent encryption in our setting. In light of our earlier empirical measurements and analytic modeling, we hypothesize that the remaining gap is due to \emph{local} evidence at low coverage, which we address next with chunk-level models.

\subsection{Chunk-Level CNNs Operation}
\subsubsection{Rationale and Overview}
Global, whole-file histograms dilute the weak signal left by intermittent encryption at low coverage: most bytes remain plaintext, so averaged statistics wash out the evidence. Our empirical atlas (Sec.~\ref{datas1}) shows this directly (flat curves for near-uniform types at low $\alpha$), and the mixture model (Sec.~\ref{sec:stat-model}) explains why: detectability is driven by small pockets where plaintext structure still dominates locally. 

Our core idea is to localize the decision. We partition each file into fixed-size byte chunks, encode each chunk as a $16{\times}16$ histogram image, and run a CNN per chunk to score whether it contains encrypted content. File-level decisions then aggregate chunk scores to trigger early while controlling false alarms.

The closest prior work that implemented some form of chunk-based approach were primarily header-centric detectors (e.g.,~\cite{e24101503,kim2022byte}) that target the file header because it has well-specified fields and magic values for each type. Under \emph{full-file} encryption the header is inevitably modified, so anomalies there are strong signals. Intermittent encryption breaks these assumptions: (i) there is no guarantee \emph{where} encryption lands or \emph{how much} is encrypted (smart/dot patterns can bypass headers entirely), and (ii) away from the header, most regions lack standardized structure to compare against, so “known-field” checks do not transfer. Our approach therefore scans \emph{across the whole file} with chunk-level histogram features, avoiding reliance on type-specific header semantics while remaining sensitive to small, arbitrarily placed encrypted pockets.

\subsubsection{Our Approach}
We make a file-level decision by aggregating \emph{chunk-level} CNN predictions via the following process:

\textbf{a) Chunking.} Let the file length be $N$ bytes. We split it into non-overlapping chunks of size $L=10{,}240$ bytes (10\,KB), giving
\(
K \;=\; \left\lceil \frac{N}{L} \right\rceil
\)
chunks indexed by $j=1,\ldots,K$ (the last chunk may be shorter).

\textbf{b) Per-chunk inference (by the CNN).} For each chunk $j$, we build its $16{\times}16$ histogram image $H_j$ (Sec.~\ref{sec:encoding}) and pass it through the CNN, which outputs a binary label
\[
\hat{y}_j \in \{0,1\} \quad\text{(1 = encrypted, 0 = not encrypted)}
\]
via the network’s own decision rule (argmax over two logits). We do \emph{not} introduce an additional, hand-set threshold at the chunk level.

\textbf{c) File-level aggregation.} Let
\(
p \;=\; \frac{1}{K}\sum_{j=1}^{K} \hat{y}_j
\)
be the fraction of chunks the CNN labeled as encrypted. We label the file as encrypted if
\(
p \;\ge\; t_{\mathrm{file}}(F),
\)
where $t_{\mathrm{file}}(F)\in[0,1]$ is a \emph{file-type–specific} threshold (for type $F$) chosen on the validation set (grid over $[0,1]$). In deployments that target a fixed false-positive rate, $t_{\mathrm{file}}(F)$ is set to meet the desired FPR.

As discussed earlier, the type-specific thresholds are motivated by native randomness differing by format: compressed/container types (e.g., DOCX, PDF) exhibit higher baseline entropy than lossless/structured types (e.g., BMP, XLS). As a result, the CNN tends to flag a larger fraction of chunks in high-entropy families even when no attack is present. Using a higher file-level aggregation threshold $t_{\mathrm{file}}(F)$ reduces false positives for these families, while a lower $t_{\mathrm{file}}(F)$ preserves sensitivity on more structured types. We select $t_{\mathrm{file}}(F)$ on the validation set (grid over $[0,1]$). This adaptive approach significantly improves detection accuracy across diverse file types and encryption patterns as shown in the following subsections.

\subsection{Chunk-Level Results}
\subsubsection{Performance Across Modes of Intermittent Encryption}
We first compare the chunk-level model across the four patterns observed in ransomware families (Table~\ref{encryption-schemes-used}, Page \pageref{encryption-schemes-used}): \emph{Adaptive}, \emph{Dot Pattern}, \emph{Head Only}, and \emph{Hybrid}. For this analysis we keep the training recipe, chunk size (10\,KB), and file-type aggregation thresholds fixed; only the encryption mode varies. For the Dot pattern, we use $N=64KB$ and $X=\alpha$  (the encryption percentage). For the Head Only pattern we use $X=\alpha$. Since the Hybrid pattern combines the Dot and Head Only patterns, it inherits these exact parameter values. In the Adaptive pattern, we used the Hybrid pattern for files greater than 10 $MB$ and full encryption for smaller files. 

\begin{figure}
\centering
    \includegraphics[width=0.50\columnwidth]{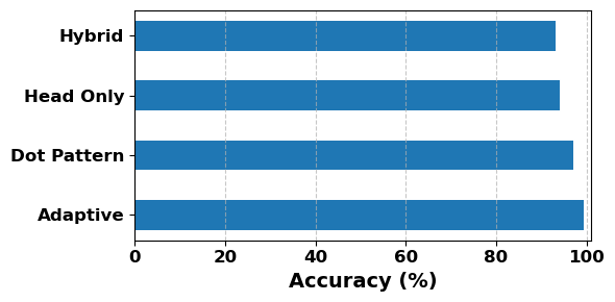}
    \caption{Average detection accuracy for each encryption scheme category}
    \label{encryption-schemes}
\end{figure}

Figure~\ref{encryption-schemes} shows that accuracy \textbf{is uniformly high across modes, with the Adaptive mode being slightly higher than the others}. This is likely due to the Adaptive mode doing 100\% encryption on the smaller files, potentially making them easier to spot than if they had only seen partial encryption as done by the other modes. Overall, for the settings described in the previous paragraph, the chunk-level detector performs very well across all encryption modes.

\begin{figure}
    \centering

    \subfloat[Dot Pattern 25\% Encryption\label{fig:dp25}]{
        \includegraphics[width=0.24\textwidth]{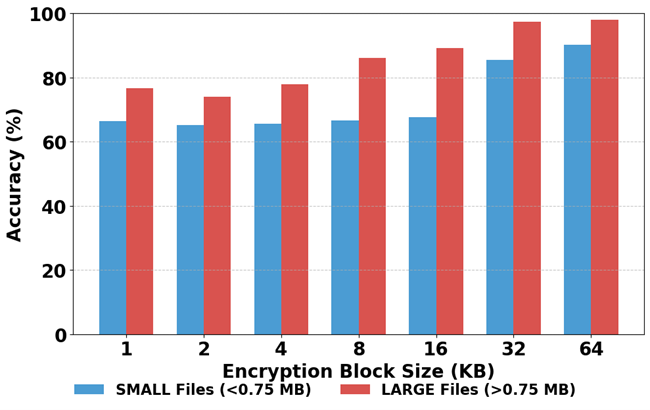}
    }
    \hfill
    \subfloat[Dot Pattern 75\% Encryption\label{fig:dp75}]{
        \includegraphics[width=0.22\textwidth]{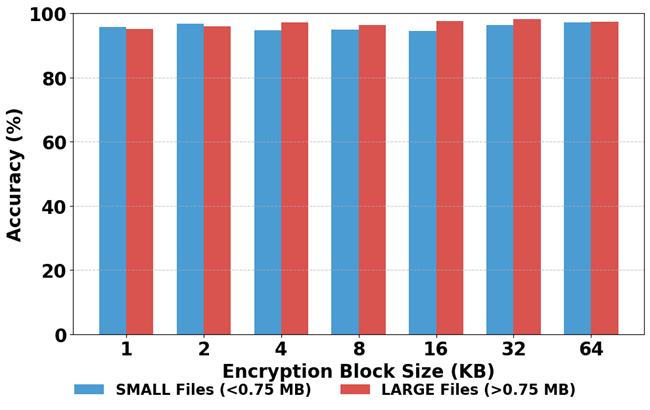}
    }

    \par\vspace{0.75em}

    \subfloat[Hybrid 25\% Encryption\label{fig:asp25}]{
        \includegraphics[width=0.22\textwidth]{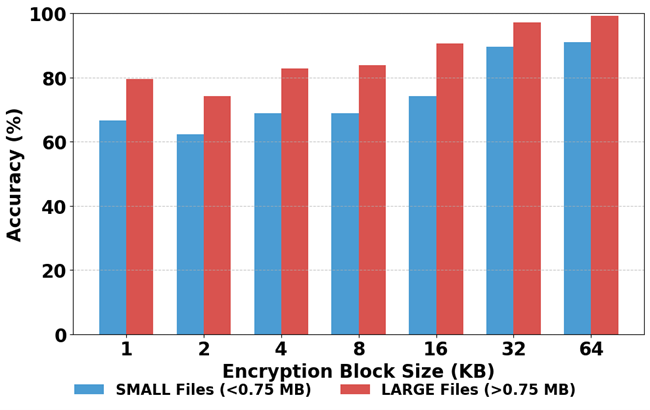}
    }
    \hfill
    \subfloat[Hybrid 75\% Encryption\label{fig:asp75}]{
        \includegraphics[width=0.24\textwidth]{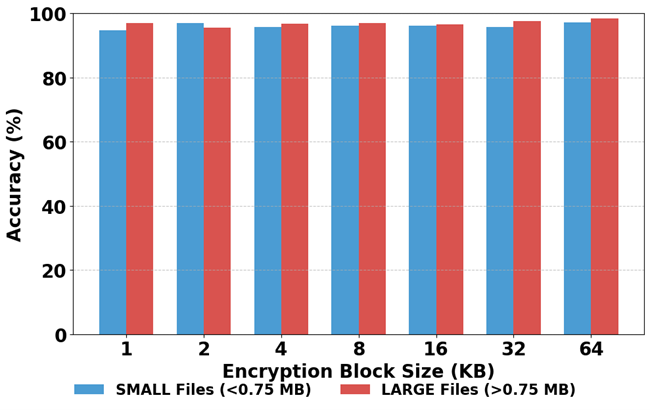}
    }

    \caption{Accuracy versus encryption block size for small and large files.}
    \label{fig:dp-asp-compare}
\end{figure}

Figure \ref{fig:dp-asp-compare} shows the results when we conducted a sensitivity analysis on two key parameters: $N$ and $\alpha$. The values $\alpha$=25\% and 75\% respectively capture cases of low and high encryption coverage. The x-axis of the plots captures the values of $N$, with the largest one (64 KB) being the one used for the earlier described result in Figure \ref{encryption-schemes}. At high encryption coverage (Figures \ref{fig:dp75} and \ref{fig:asp75}),  accuracy is so close to 100\% for all values of $N$. At low encryption coverage however (Figures \ref{fig:dp25} and \ref{fig:asp25}), smaller values of $N$ show clear decline in performance, with the smaller files always performing worse than the larger ones. 

This is likely because, at low encryption coverage, each segment of the Dot pattern (or Hybrid pattern) sees a very small amount of encryption which might not generate enough evidence for occurrence of encryption. With 75\% encryption on the other hand, each encrypted file will have a lot of evidence to confirm encryption of the segment.

% \begin{figure*}[t]
%   \centering

%   \begin{subfigure}[t]{0.45\textwidth}
%     \centering
%     \includegraphics[width=\linewidth]{Figures/dp_25.png}
%     \caption{Dot Pattern 25\% Encryption}
%     \label{fig:dp25}
%   \end{subfigure}
%   \hfill
%   \begin{subfigure}[t]{0.45\textwidth}
%     \centering
%     \includegraphics[width=\linewidth]{Figures/dp_75.png}
%     \caption{Dot Pattern 75\% Encryption}
%     \label{fig:dp75}
%   \end{subfigure}

%   \vspace{0.75em}

%   \begin{subfigure}[t]{0.45\textwidth}
%     \centering
%     \includegraphics[width=\linewidth]{Figures/asp_25.png}
%     \caption{Hybrid 25\% Encryption}
%     \label{fig:asp25}
%   \end{subfigure}
%   \hfill
%   \begin{subfigure}[t]{0.45\textwidth}
%     \centering
%     \includegraphics[width=\linewidth]{Figures/asp_75.png}
%     \caption{Hybrid 75\% Encryption}
%     \label{fig:asp75}
%   \end{subfigure}

%   \caption{Accuracy vs.\ encryption block size for SMALL and LARGE files.}
%   \label{fig:dp-asp-compare}
% \end{figure*}

\begin{table}[h]
\centering
\scriptsize

\caption{Accuracy Across Encryption Coverage by Filetype}
\label{tab:accuracy1}
\begin{tabular}{l | r r r r r r}
\toprule
\textbf{Filetype} & \multicolumn{6}{c}{\textbf{Encryption Coverage (\%)}} \\
\cmidrule(l){2-7}
                  & \textbf{10} & \textbf{25} & \textbf{50} & \textbf{75} & \textbf{90} & \textbf{100} \\
\midrule
BMP   & 98.50\% & 98.50\% & 98.50\% & 98.50\% & 98.50\% & 98.50\% \\
DOC   & 89.00\% & 99.00\% & 99.00\% & 99.00\% & 99.00\% & 99.00\% \\
DOCX  & 73.50\% & 82.00\% & 97.50\% & 97.50\% & 97.50\% & 97.50\% \\
JPG   & 100.00\% & 100.00\% & 100.00\% & 100.00\% & 100.00\% & 100.00\% \\
MP3   & 99.50\% & 99.50\% & 99.50\% & 99.50\% & 99.50\% & 99.50\% \\
MP4   & 100.00\% & 100.00\% & 100.00\% & 100.00\% & 100.00\% & 100.00\% \\
PDF   & 88.00\% & 93.00\% & 93.00\% & 93.00\% & 93.00\% & 93.00\% \\
PNG   & 99.50\% & 99.50\% & 99.50\% & 99.50\% & 99.50\% & 99.50\% \\
PPT   & 84.50\% & 97.00\% & 97.00\% & 97.00\% & 97.00\% & 97.00\% \\
TXT   & 94.68\% & 94.68\% & 94.68\% & 94.68\% & 100.00\% & 94.68\% \\
XLS   & 91.00\% & 94.00\% & 94.00\% & 94.00\% & 94.00\% & 94.00\% \\
\bottomrule
\end{tabular}
\end{table}

\subsubsection{Performance Across File Types and Coverage}
Table~\ref{tab:accuracy1} reports file-level accuracy for the chunk model by file type and encryption coverage ($10$--$100\%$).

\textbf{High coverages are handled very well by the classifier.}
From $25\%$ upward, \emph{most} types exceed $\sim\!95\%$ accuracy (many at $99$--$100\%$), clearly outperforming the earlier whole-file/GAN baselines. By $50\%$ coverage, the model is near-saturated for every type.

\textbf{The only challenging regime is $10\%$ coverage.}
Here we see spread across types: \texttt{docx} is lowest (73.50\%), with \texttt{ppt} (84.50\%), and \texttt{pdf} (88.00\%) also below the top cluster. Two plausible, data-driven reasons:
% \textbf{The only challenging regime is $10\%$ coverage.}
% Here we see spread across types: \texttt{docx} is lowest (73.50\%), with \texttt{xls} (84.25\%), \texttt{ppt} (80.50\%), and \texttt{pdf} (88.00\%) also below the top cluster. Two plausible, data-driven reasons:

- \emph{Container/compressed formats} (e.g., \texttt{docx}) are ZIP-based and already high-entropy; a small encrypted fraction adds little separable structure at chunk scale, so distinguishing “compressed vs.\ encrypted” is hardest at low coverage.

- \emph{Chunk/aggregation effects} can depress accuracy for types with fewer chunks per file (smaller $K$): with only a handful of chunks, $10\%$ coverage may touch just one chunk, and the file-level proportion may fall below the type threshold $t_{\mathrm{file}}(F)$.

The \texttt{xls} dip at $10\%$ is notable given its structured (BIFF) format; we hypothesize a combination of smaller average file sizes (lower $K$) and format-specific structure that makes some plaintext chunks resemble our encrypted class at very low signal. We leave a targeted exploration (varying chunk length/stride and using a top-$m$ aggregator) to future work.

In general, chunk-level inference delivers uniformly high accuracy from $25\%$ coverage upward across all file types. The remaining frontier is the ultra–low-coverage regime ($\sim\!10\%$), especially for container/compressed families and smaller files, where chunk count and compressed-vs-encrypted separability dominate. 

It is noteworthy that we include the 10\% setting mostly for completeness, but such extreme under-encryption is unlikely in practice: at $\alpha\!\approx\!0.10$ many formats remain partially usable or recoverable (e.g., ZIP-containered \texttt{docx}/\texttt{pptx} can still expose content if central metadata is intact; media stays playable with limited degradation). Ransomware operators typically choose higher fractions to guarantee denial of access. Accordingly, our threat emphasis is on $\alpha\!\ge\!25\%$; the 10\% band serves more as a stress test for detector sensitivity rather than a realistic operating point.

\subsubsection{Wrapping up: Comparing Global Performance across Training Strategies}

\begin{figure}
    \centering
    \includegraphics[width=0.5\columnwidth]{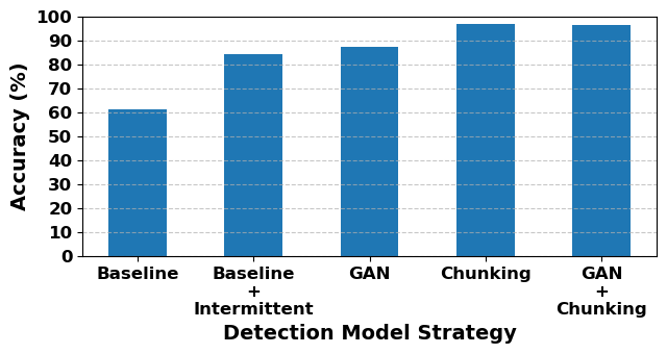}
    \caption{Comparison of model mean accuracy across training strategies}
    \label{model-strategy}
\end{figure}

Figure \ref{model-strategy} summarises overall detection accuracy for five training pipelines.
The \textbf{Baseline} model—trained only on fully encrypted and plaintext files—achieves the lowest score ($\approx 61.5\%$), revealing how conventional whole‑file approaches falter against partial encryption.
Augmenting the data set with \textbf{intermittently encrypted} samples (\textbf{Baseline+Intermittent}) raises accuracy to roughly 84.4\%, but the classifier still mislabels many lightly encrypted regions.
Adding \textbf{GAN‑generated} synthetic files lifts performance further to about 87.34\%, showing that high‑entropy augmentations help, yet the improvement is modest.

The decisive gain comes from the \textbf{Chunking} strategy: by slicing every file into fixed 10KB blocks before classification, the model reaches \textbf{96.54\%} accuracy.

Combining synthetic data with chunking (GAN+Chunking) offers no meaningful extra benefit, plateauing at roughly the same level as chunking alone.

In short, the experiment shows that \emph{granularity} is the dominant factor: moving from whole‑file to block‑level analysis eliminates most blind spots, while data augmentation provides smaller, incremental gains. 

\section{Conclusion}
We presented three complementary contributions: (i) a systematic, multi–file-type empirical atlas that traces how byte-level structure evolves under intermittent encryption; (ii) a compact analytic model that specializes KL-based bounds to this setting and yields per-family detectability ceilings via a single constant $c_F^2$; and (iii) a practical detection pipeline that turns these insights into high performance, with a chunk-level CNN achieving $\approx97\%$ accuracy in our best configuration.

%Global (whole-file) histogram cues erode quickly on near-uniform media, while structured formats remain detectable longer—an asymmetry our mixture model explains and quantifies. Training on intermittent examples closes most of the gap, but the decisive gain comes from localizing the decision: evaluating 10\,KB chunks and aggregating at file level. From $25\%$ coverage upward, accuracy is uniformly high across types. The remaining hard corner is the ultra-low-coverage regime ($\sim\!10\%$), especially for container/compressed families (e.g., DOCX) and smaller files, where compressed-vs-encrypted separability and low chunk counts constrain performance. Operationally, however, attackers rarely stay this low: very small encrypted fractions risk partial file recovery, so practical campaigns tend to use higher coverage or target critical regions. Thus our detector performs strongest in the same operating window attackers prefer.

Our approach inspects the post-attack artefact (the modified file), so its signal persists regardless of how the encryption was orchestrated (e.g., throttling, API unhooking, etc.). It is therefore complementary to behavioural defences that aim to stop encryption early. Combined, they form a layered strategy: runtime monitors to cut off damage quickly; content analysis to catch stealthy, intermittent campaigns and support retrospective scans of endpoints and backups.

%The empirical trajectories and the $c_F^2$ constants inform file-type–aware thresholding and calibration for any histogram-driven classifier. More broadly, the atlas and model support incident-response triage (estimating likely coverage to prioritise restores), forensic analysis (distinguishing compression from encryption drift), backup integrity audits, and data-loss investigations where partial ciphertext may be present.

%Our study uses fixed $16{\times}16$ byte-histogram images and a single chunk size (10\,KB); very small files and formats with few chunks are intrinsically harder at low coverage. The dataset, while diverse, cannot span every enterprise corpus. Finally, GAN augmentation did not consistently help, suggesting that matching real compressed–vs–encrypted variability remains challenging.

%Several directions look promising: multi-scale chunking and overlap/stride tuning to boost sensitivity at ultra-low coverage; calibrated, type-aware aggregation rules (e.g., top-$m$ voting or Bayesian pooling); self-supervised pretraining on unlabeled corpora to improve per-chunk features; joint content+behaviour ensembles for earlier, higher-confidence alarms; and rigorous in-the-wild evaluations across organisations, including cost/latency profiling and policy tuning to fixed false-positive budgets.

By coupling a first-principles model with a localized learning pipeline, we both explain \emph{why} partial encryption evades traditional detectors and demonstrate \emph{how} to close that gap in practice. The resulting chunk-level CNN provides a strong, practical baseline for defenders and a clear benchmark for future systems.% to beat.

% \bibliographystyle{ACM-Reference-Format}
% \bibliography{main}
\bibliographystyle{IEEEtran}
\bibliography{main}

% \bibliographystyle{ACM-Reference-Format}
% \bibliography{codaspy_main_fixed}

%%%%%%%%%%%%%%%%%%%%%%%%%%%%%%%%%%
\newpage
\section{Appendix A: Detailed proof of Proposition~\ref{thm:family-threshold}}
\label{append}
This appendix gives the full proof of Proposition~\ref{thm:family-threshold}, specializing a well-known KL upper bound \cite{DragomirGluscevic2000}, to our intermittent-encryption problem. Our contribution is to rewrite the ceiling in terms of a directly measurable family constant $c_F^{2}$ and the encrypted fraction $\alpha$, yielding file-type–specific detectability ceilings and simple rules for thresholding/normalization across types. In practice, we estimate $c_F^{2}$ empirically from cleartext corpora and report medians with 95\% CIs in Table~\ref{tf-values}; these constants plug directly into the ceiling and the operating rules discussed in Section~\ref{sec:stat-model} (e.g., $\alpha^\star$ and KL normalization).

The \emph{tightness} of the bound relies on a \emph{small-leak} regime, $\max_b|\varepsilon_b(\alpha)|<1$. This condition has a natural interpretation in ransomware settings: (i) attackers typically operate at high coverage ($\alpha\!\approx\!1$) to hinder recovery while leaving just enough plaintext to evade detectors, and (ii) for near-uniform/compressed families $\Delta(b)$ is intrinsically small, so the condition holds across a wide range of $\alpha$. Outside this regime the inequality still holds but may be conservative; our use focuses on the attacker-relevant high-coverage zone where it is most informative. For completeness we present the full derivation below.

Expounding on the abstraction of the intermittent encryption process formulated in Section \ref{formulate}, define the deviation,
\[
  \Delta(b) := P_{\text{orig},F}(b) - U(b).
\]
The mixture density in Equation \ref{eq:pmix} (see Section \ref{formulate}), can be expressed in terms of the deviation,
\[
  P_{\text{mix}}(\alpha,b) = U(b) + (1-\alpha)\,\Delta(b).
\]

Because both $P_{\text{orig},F}$ and $U$ are probability distributions, they
each sum to 1 across the entire domain, hence:
\[
  \sum_{b=0}^{255} \Delta(b)
    = \sum_{b}P_{\text{orig},F}(b)-\sum_{b}U(b)
    = 1-1 = 0.
\]

Substituting $U(b)=\tfrac{1}{256}$ into the mixture definition, $P_{\text{mix}}(\alpha,b)$, and
factoring out the common $\frac{1}{256}$ term gives,
\[
  P_{\text{mix}}(\alpha,b)
  = \frac{1}{256}\Bigl[\,1 + 256(1-\alpha)\,\Delta(b)\Bigr].
\]
Introducing the substitution,
\[
  \varepsilon_b(\alpha) := 256(1-\alpha)\,\Delta(b),
\]
gives,
\[
  P_{\text{mix}}(\alpha,b)
  = \frac{1}{256}\,\bigl[1+\varepsilon_b(\alpha)\bigr].
\]

%------------------------------------------------------------
The KL divergence in \emph{bits} is
\[
  D_{\mathrm{KL}}\!\bigl(P_{\mathrm{mix}}(\alpha)\,\|\,U\bigr)
   =\sum_{b} P_{\text{mix}}(\alpha,b)
      \log_{2}\!\frac{P_{\text{mix}}(\alpha,b)}{U(b)}.
\]
Since $U(b)=1/256$,
\[
  \frac{P_{\text{mix}}(\alpha,b)}{U(b)}
    = 256\,P_{\text{mix}}(\alpha,b)
    = 1+\varepsilon_b(\alpha).
\]
Therefore,
\begin{equation}
     D_{\mathrm{KL}}
   = \sum_{b}\frac{1}{256}\bigl[1+\varepsilon_b(\alpha)\bigr]\,
       \log_{2}\!\bigl(1+\varepsilon_b(\alpha)\bigr)
       \label{eq:kl}
\end{equation}

Having formulated $D_{\mathrm{KL}}$ in the context of our problem space, we now proceed to find its upper bound, which will enable us reason about detection thresholds. 

Because $\sum_b \Delta(b)=0$, we have,
\begin{equation*}
\sum_{b}\varepsilon_b(\alpha)
 = 256(1-\alpha)\sum_b \Delta(b)
 = 0.
\end{equation*}
Also,
\begin{equation*}
\sum_{b}\varepsilon_b(\alpha)^{2}
 = 256^{2}(1-\alpha)^{2}\sum_{b}\Delta(b)^{2}
 = 256^{2}(1-\alpha)^{2}c_F^{2},
\end{equation*}
where $c_F^{2}=\sum_{b}\Delta(b)^{2}=\|\Delta\|_{2}^{2}$.

For $|x|<1$, $\log_{2}(1+x)\le x/\ln 2$.  Hence, given,
$\max_b |\varepsilon_b(\alpha)| < 1$, we can upper bound $D_{\mathrm{KL}}$ as below (linear term vanishes), 
\begin{align*}
D_{\mathrm{KL}}\bigl(P_{\text{mix}}(\alpha)\,\|\,U\bigr)
 &\le \sum_{b}\frac{1}{256}\bigl[1+\varepsilon_b(\alpha)\bigr]
                 \frac{\varepsilon_b(\alpha)}{\ln 2}
      \notag\\
 &= \frac{1}{256\,\ln 2}\sum_{b}\varepsilon_b(\alpha)^{2}.
\end{align*}

% Note that the condition $\max_b |\varepsilon_b(\alpha)| < 1$ is tight in two practically relevant cases.

% \textbf{(i) High‑encryption regime:} $\alpha$ is close to 1, regardless of file family.
% Whenever feasible, attackers would prefer to operate in this zone because it makes format‑aware file recovery tools (e.g., White Phoenix \cite{BleepingComputer2023WhitePhoenix}) far less effective. The attacker's goal is to leave enough of the file unencrypted to just evade detection, while still encrypting the bulk of the file to thwart reconstruction. 

% \textbf{(ii) Near‑uniform families:} for highly compressed/encoded media (DOCx, JPEG, etc.),
% $\Delta(b)$ is intrinsically tiny, so $|\varepsilon_b(\alpha)|<1$ holds across a wide range
% of $\alpha$.\\

Now substituting for, $\sum_{b}\varepsilon_b(\alpha)^{2}=256^{2}(1-\alpha)^{2}c_F^{2}$:
\begin{align}
D_{\mathrm{KL}}\bigl(P_{\text{mix}}(\alpha)\,\|\,U\bigr)
 &\le \frac{1}{256\,\ln 2}\,256^{2}(1-\alpha)^{2}c_F^{2}
 \notag\\
 &= \frac{256}{\ln 2}\,c_F^{2}(1-\alpha)^{2}.
\label{eqfinal}
\end{align}

Equation~\eqref{eqfinal} highlights that for a fixed file family
(which means $c_F^{2}$ is restricted within some range), KL can never exceed, $\frac{256}{\ln 2}\,c_F^{2}(1-\alpha)^{2}$.
Thus any classification threshold,  $\tau_F$ set above that value is information‑theoretically guaranteed to not detect the attack. Thresholds below it may
or may not work depending on sampling noise and the detector.

%------------------------------------------------------------
% \subsection*{A.7  Conclusion}

% Equation (A.8) is the quadratic approximation
% Eq.~\eqref{eq:kl-quad} in Theorem~\ref{thm:kl-escape}.
% The remainder is dominated by the cubic term, so the approximation is
% accurate whenever
% \[
%   |\varepsilon_b(\alpha)| < 1\quad\forall\,b
%   \;\;\Longleftrightarrow\;\;
%   (1-\alpha) < \frac{1}{256\,\max_b|\Delta(b)|}.
% \]
% For structured files this covers leaks below a few percent; for
% near‑uniform lossy files it covers essentially the whole \(\alpha\)-range.

\hfill\(\square\)

% \section{Appendix A: Detailed proof of Proposition~\ref{thm:family-threshold}}
%%%%%%%%%%%%%%%%%%%%%%%%%%%%%%%%%%%
% \subsection{Entropy}

\begin{table}[H]
\tiny
\centering

\label{tab:entropy_medians}
\caption{The median of bytewise \textit{entropy} (bits/byte) across file types as the encrypted fraction \(\alpha\) increases. Medians rise monotonically with \(\alpha\) for all types and converge near the 8\,bits/byte ceiling by \(\alpha=1\). Already-compressed media (PNG, JPG, MP3, MP4) start \(\approx\)7.98–7.99 and change minimally, while semi-compressed containers (PDF, DOCX) are high at baseline and saturate by \(\alpha\!\approx\!0.7\). Structured/uncompressed formats (txt, DOC, PPT, BMP) begin lower and show the largest gains (e.g., txt median 5.24\(\rightarrow\)7.99).}
\textbf{\begin{tabular}{l|cccccccccc}
\toprule
Filetype & $\alpha{=}0.1$ & 0.2 & 0.3 & 0.4 & 0.5 & 0.6 & 0.7 & 0.8 & 0.9 & 1.0 \\
\midrule
txt   & {5.24} & 5.77 & 6.22 & 6.62 & 6.97 & 7.28 & 7.55 & 7.76 & 7.92 & 7.99 \\
DOC   & {6.57} & 6.85 & 7.06 & 7.16 & 7.24 & 7.36 & 7.49 & 7.65 & 7.83 & 8.00 \\
PPT   & {7.68} & 7.70 & 7.72 & 7.74 & 7.76 & 7.78 & 7.81 & 7.85 & 7.92 & 8.00 \\
BMP   & {7.50} & 7.56 & 7.61 & 7.65 & 7.70 & 7.76 & 7.83 & 7.90 & 7.97 & 8.00 \\
DOCX  & {7.92} & 7.93 & 7.94 & 7.95 & 7.96 & 7.97 & 7.98 & 7.98 & 7.99 & 8.00 \\
PDF   & {7.94} & 7.94 & 7.95 & 7.95 & 7.95 & 7.96 & 7.96 & 7.96 & 7.97 & 8.00 \\
PNG   & {7.99} & 7.99 & 7.99 & 7.99 & 7.99 & 8.00 & 8.00 & 8.00 & 8.00 & 8.00 \\
MP4   & {7.97} & 7.97 & 7.98 & 7.98 & 7.99 & 7.99 & 7.99 & 8.00 & 8.00 & 8.00 \\
MP3   & {7.97} & 7.98 & 7.98 & 7.99 & 7.99 & 7.99 & 8.00 & 8.00 & 8.00 & 8.00 \\
JPG   & {7.98} & 7.98 & 7.99 & 7.99 & 7.99 & 8.00 & 8.00 & 8.00 & 8.00 & 8.00 \\
\bottomrule
\end{tabular}}
\end{table}

%%%%%%%%%%%%%%%%%%%%%%%%%%%%%%%%%
% \subsection{variance}

\begin{table}[H]
\tiny
\centering
\setlength{\tabcolsep}{5pt}
\caption{The median of bytewise variance (\(\times 10^{-6}\)) across file types as the encrypted fraction \(\alpha\) increases. Medians fall monotonically with \(\alpha\) for all types and approach \(\approx 0\) by \(\alpha=1\), reflecting progressive flattening of the byte histogram. The steepest declines occur for structured/uncompressed formats (txt, DOC, PPT, BMP): e.g., txt median 186.06\(\rightarrow\)0.15, while DOC’s median collapses from \(\approx 232.93\) at \(\alpha=0.1\) to \(\approx 0.02\) at \(\alpha=1.0\). Semi-compressed containers (PDF, DOCX) start low and smoothly decay toward \(\approx 0\), and already-compressed media (PNG, JPG, MP3, MP4) are low at baseline and rapidly plateau near zero by \(\alpha\gtrsim 0.8\).}
\label{tab:variance_medians}
% \resizebox{\columnwidth}{!}{%
\textbf{\begin{tabular}{l|cccccccccc}
\toprule
\rotatebox{90}{Filetype}& $\alpha{=}0.1$ & 0.2 & 0.3 & 0.4 & 0.5 & 0.6 & 0.7 & 0.8 & 0.9 & 1.0 \\
\midrule
txt   & {186.06} & 147.12 & 112.74 & 82.94 & 57.49 & 36.80 & 20.80 & 9.30 & 2.44 & 0.15 \\
DOC   & {232.93} & 162.20 & 122.33 & 107.21 & 93.86 & 75.26 & 58.32 & 34.57 & 12.52 & 0.02 \\
PPT   & {23.81}  & 21.98 & 20.83 & 19.33 & 17.32 & 15.49 & 13.17 & 9.15 & 4.17 & 0.00 \\
BMP   & {19.63}  & 17.32 & 15.45 & 13.87 & 11.79 & 9.36 & 6.16 & 3.30 & 0.96 & 0.00 \\
DOCX  & {3.09}   & 2.71  & 2.34  & 1.94  & 1.52  & 1.12  & 0.78  & 0.50  & 0.28 & 0.02 \\
PDF   & {2.41}   & 2.23  & 2.15  & 2.05  & 1.90  & 1.74  & 1.57  & 1.36  & 1.05 & 0.02 \\
PNG   & {0.29}   & 0.25  & 0.20  & 0.16  & 0.11  & 0.08  & 0.05  & 0.02  & 0.01 & 0.00 \\
MP4   & {0.80}   & 0.66  & 0.53  & 0.42  & 0.32  & 0.23  & 0.16  & 0.10  & 0.06 & 0.00 \\
MP3   & {0.67}   & 0.53  & 0.41  & 0.30  & 0.21  & 0.14  & 0.08  & 0.04  & 0.01 & 0.00 \\
JPG   & {0.41}   & 0.34  & 0.27  & 0.21  & 0.15  & 0.10  & 0.06  & 0.03  & 0.01 & 0.00 \\
\bottomrule
\end{tabular}}
\end{table}

% \subsection{skewness}
%%%%%%%%%%%%%%%%%%%%%%%%%%%%%%%%%%
\begin{table}[H]
\tiny
\centering
\setlength{\tabcolsep}{5pt}
\setlength{\tabcolsep}{5pt}
\caption{The median of \textit{skewness} across file types as the encrypted fraction \(\alpha\) increases. Structured and uncompressed types such as txt, DOC, PPT, and BMP keep high and nearly constant skewness through partial encryption, then collapse to near zero at \(\alpha=1.0\). Compressed media such as PNG, JPG, and MP3 start low and drift toward zero with larger \(\alpha\). Container and compound formats such as MP4, DOCX, and PDF often rise modestly under partial encryption and then converge to near zero at full encryption. Small negative medians at \(\alpha=1.0\) reflect estimation noise around a symmetric distribution.}
\label{tab:skewness_medians}

\textbf{\begin{tabular}{l|cccccccccc}
\toprule
\rotatebox{90}{Filetype} & $\alpha{=}0.1$ & 0.2 & 0.3 & 0.4 & 0.5 & 0.6 & 0.7 & 0.8 & 0.9 & 1.0 \\
\midrule
txt   & {6.01} & 6.01 & 6.00 & 5.99 & 5.99 & 5.98 & 5.95 & 5.85 & 5.40 & 0.09 \\
DOC   & {15.33} & 15.34 & 15.35 & 15.37 & 15.39 & 15.42 & 15.49 & 15.56 & 15.61 & 0.02 \\
PPT   & {13.21} & 13.29 & 13.35 & 13.40 & 13.41 & 13.44 & 13.49 & 13.51 & 13.43 & 0.02 \\
BMP   & {6.76} & 6.76 & 6.80 & 6.86 & 6.76 & 6.77 & 6.83 & 6.89 & 6.95 & 0.01 \\
DOCX  & {10.11} & 10.06 & 10.05 & 10.22 & 10.80 & 11.16 & 12.03 & 12.83 & 13.27 & 0.04 \\
PDF   & {9.17} & 9.50 & 9.74 & 10.00 & 10.21 & 10.37 & 10.57 & 10.76 & 11.02 & 0.03 \\
PNG   & {0.78} & 0.79 & 0.78 & 0.78 & 0.77 & 0.74 & 0.70 & 0.62 & 0.42 & 0.01 \\
MP4   & {4.17} & 4.42 & 4.72 & 5.08 & 5.63 & 6.38 & 7.52 & 9.30 & 11.73 & 0.01 \\
MP3   & {4.31} & 4.29 & 4.21 & 4.20 & 4.11 & 4.04 & 3.84 & 3.40 & 1.97 & 0.02 \\
JPG   & {1.50} & 1.36 & 1.27 & 1.23 & 1.18 & 1.16 & 1.09 & 0.95 & 0.57 & 0.01 \\
\bottomrule
\end{tabular}}
\end{table}
 
% \subsection{Euclidean distance}

\begin{table}[H]
\tiny
\centering
\setlength{\tabcolsep}{5pt}
\caption{The median of \textit{Euclidean distance} across file types as the encrypted fraction \(\alpha\) increases. Distances grow monotonically with \(\alpha\) for all types. Structured and uncompressed content such as txt, DOC, PPT, and BMP and compound types such as DOCX and PDF show rapid growth, reaching very large values by \(\alpha=1.0\). Compressed media such as PNG, JPG, MP3, and MP4 increase more gradually and remain much smaller than text and legacy office formats even at full encryption.
}
\label{tab:euclid_medians}

\textbf{\begin{tabular}{l|cccccccccc}
\toprule
\rotatebox{90}{Filetype} & $\alpha{=}0.1$ & 0.2 & 0.3 & 0.4 & 0.5 & 0.6 & 0.7 & 0.8 & 0.9 & 1.0 \\
\midrule
txt   & {23.84} & 48.16 & 72.42 & 96.69 & 120.95 & 145.39 & 169.58 & 193.70 & 217.95 & 241.52 \\
DOC   & {26.41} & 55.37 & 80.71 & 100.56 & 113.88 & 127.43 & 148.48 & 178.11 & 215.52 & 268.79 \\
PPT   & {3.28}  & 5.17  & 7.44  & 10.28 & 13.68 & 18.67 & 25.39 & 36.18 & 53.04 & 80.97 \\
BMP   & {7.16}  & 13.12 & 18.25 & 22.85 & 27.23 & 32.39 & 39.43 & 49.68 & 61.63 & 75.81 \\
DOCX  & {6.73}  & 8.21  & 10.23 & 11.97 & 13.84 & 15.30 & 18.24 & 21.38 & 24.87 & 35.63 \\
PDF   & {2.73}  & 3.94  & 5.06  & 6.14  & 6.92  & 7.79  & 8.65  & 10.21 & 12.57 & 26.93 \\
PNG   & {0.81}  & 1.57  & 2.42  & 3.30  & 4.25  & 5.22  & 6.20  & 7.16  & 8.14  & 9.13 \\
MP4   & {1.58}  & 2.97  & 4.35  & 5.76  & 7.13  & 8.55  & 9.96  & 11.35 & 12.80 & 15.85 \\
MP3   & {2.04}  & 3.45  & 4.90  & 6.35  & 7.81  & 9.28  & 10.73 & 12.16 & 13.59 & 15.05 \\
JPG   & {2.05}  & 3.39  & 4.47  & 5.45  & 6.35  & 7.28  & 8.25  & 9.25  & 10.29 & 11.41 \\
\bottomrule
\end{tabular}}

\end{table}

%%%%%%%%%%%%%%%%%%%%%%%%%%%%%%%%%%%%%%%%%%%%%
% \subsection{KL divergence}

\begin{table}[H]
\centering
\tiny
\setlength{\tabcolsep}{5pt}
\caption{The median of\textit{ KL divergence} across file types as the encrypted fraction \(\alpha\) increases. Values rise monotonically for all types. Text and legacy office content such as DOC, PPT, and BMP show the fastest growth and very large medians by full encryption. Container documents such as DOCX and PDF increase more moderately. Precompressed media such as PNG JPG MP3 and MP4 remain small and tightly clustered relative to textual and legacy office formats across all \(\alpha\).}
\label{tab:kl_medians}

\textbf{\begin{tabular}{l|cccccccccc}
\toprule
\rotatebox{90}{Filetype} & $\alpha{=}0.1$ & 0.2 & 0.3 & 0.4 & 0.5 & 0.6 & 0.7 & 0.8 & 0.9 & 1.0 \\
\midrule
txt   & {11.26} & 24.29 & 39.15 & 56.32 & 76.46 & 100.48 & 130.37 & 169.28 & 226.97 & 342.14 \\
DOC   & {2.97}  & 7.80  & 14.36 & 18.08 & 21.26 & 24.72 & 31.55 & 45.63 & 70.17 & 172.17 \\
PPT   & {0.08}  & 0.23  & 0.48  & 0.82  & 1.31  & 1.95  & 3.04  & 4.85  & 9.58  & 33.87 \\
BMP   & {1.07}  & 3.29  & 6.04  & 9.04  & 12.42 & 15.64 & 19.92 & 25.99 & 34.50 & 57.14 \\
DOCX  & {0.14}  & 0.26  & 0.41  & 0.65  & 0.98  & 1.46  & 2.13  & 3.16  & 4.51  & 10.20 \\
PDF   & {0.10}  & 0.21  & 0.35  & 0.49  & 0.65  & 0.83  & 1.01  & 1.29  & 1.96  & 7.52 \\
PNG   & {0.01}  & 0.04  & 0.10  & 0.19  & 0.32  & 0.48  & 0.69  & 0.92  & 1.19  & 1.50 \\
MP4   & {0.03}  & 0.12  & 0.26  & 0.46  & 0.72  & 1.04  & 1.42  & 1.87  & 2.43  & 3.76 \\
MP3   & {0.05}  & 0.14  & 0.30  & 0.52  & 0.80  & 1.15  & 1.58  & 2.09  & 2.71  & 3.44 \\
JPG   & {0.07}  & 0.19  & 0.34  & 0.49  & 0.67  & 0.88  & 1.13  & 1.44  & 1.80  & 2.26 \\
\bottomrule
\end{tabular}}

\end{table}
% \subsection{total variation}

\begin{table}[H]
\centering
\tiny
\setlength{\tabcolsep}{5pt}
\caption{The median of \textit{total variation} across file types as the encrypted fraction \(\alpha\) increases. All types show a steady monotonic rise.Text grows to the largest values by full encryption. Legacy office and bitmap formats DOC, PPT, and BMP rise sharply at high \(\alpha\). Container documents DOCX and PDF increase more moderately. Precompressed media PNG JPG MP3 MP4 remain lower and more tightly clustered across all \(\alpha\) although they still climb with \(\alpha\).}
\label{tab:total_variation_medians}

\textbf{\begin{tabular}{l|cccccccccc}
\toprule
\rotatebox{90}{Filetype} & $\alpha{=}0.1$ & 0.2 & 0.3 & 0.4 & 0.5 & 0.6 s& 0.7 & 0.8 & 0.9 & 1.0 \\
\midrule
txt   & {8.38}  & 16.79 & 25.22 & 33.65 & 42.08 & 50.52 & 58.95 & 67.37 & 75.77 & 84.15 \\
DOC   & {7.22}  & 13.56 & 18.60 & 20.63 & 22.19 & 23.87 & 26.97 & 31.24 & 36.50 & 43.21 \\
PPT   & {1.14}  & 1.93  & 2.76  & 3.66  & 4.61  & 5.55  & 6.93  & 8.66  & 11.13 & 15.64 \\
BMP   & {3.79}  & 7.06  & 9.85  & 12.30 & 14.49 & 16.63 & 19.22 & 22.15 & 25.78 & 29.80 \\
DOCX  & {1.14}  & 1.71  & 2.29  & 2.99  & 3.79  & 4.68  & 5.72  & 6.82  & 7.71  & 8.70 \\
PDF   & {1.31}  & 1.93  & 2.43  & 2.86  & 3.32  & 3.75  & 4.15  & 4.73  & 5.50  & 8.51 \\
PNG   & {0.51}  & 0.98  & 1.51  & 2.06  & 2.64  & 3.23  & 3.85  & 4.45  & 5.06  & 5.65 \\
MP4   & {0.85}  & 1.62  & 2.39  & 3.16  & 3.95  & 4.72  & 5.50  & 6.26  & 7.07  & 8.05 \\
MP3   & {0.85}  & 1.58  & 2.30  & 3.03  & 3.75  & 4.48  & 5.21  & 5.94  & 6.65  & 7.40 \\
JPG   & {1.16}  & 1.96  & 2.60  & 3.16  & 3.69  & 4.20  & 4.76  & 5.37  & 5.99  & 6.67 \\
\bottomrule
\end{tabular}}

\end{table}

\end{document}